\documentclass[11pt]{article}

\usepackage[final]{acl}

\usepackage{times}
\usepackage{latexsym}
\usepackage{booktabs}
\usepackage{pifont}
\usepackage{xurl}
\usepackage{siunitx}
\usepackage{amsmath}   
\usepackage{amssymb}   
\usepackage{tcolorbox}
\usepackage{xcolor}
\usepackage{listings}
\tcbuselibrary{breakable,skins,listings}
\usepackage{fancyvrb}
\usepackage{fvextra}
\usepackage{graphicx}
\usepackage{cuted}

\usepackage[T1]{fontenc}

\usepackage[utf8]{inputenc}

\usepackage{microtype}

\usepackage{inconsolata}

\usepackage{graphicx}

\title{LongDA: Benchmarking LLM Agents for Long-Document \\ Data Analysis}

\author{
 \textbf{Yiyang Li\textsuperscript{1}},
 \textbf{Zheyuan Zhang\textsuperscript{1}},
 \textbf{Tianyi Ma\textsuperscript{1}},
 \textbf{Zehong Wang\textsuperscript{1}},
\\
 \textbf{Keerthiram Murugesan\textsuperscript{2}},
 \textbf{Chuxu Zhang\textsuperscript{3}},
 \textbf{Yanfang Ye\textsuperscript{1\textdagger}}
\\
 \textsuperscript{1}University of Notre Dame,
 \textsuperscript{2}IBM Research,
 \textsuperscript{3}University of Connecticut\\
\textsuperscript{\textdagger}Corresponding Author \quad \texttt{\{yli62, yye7\}@nd.edu}
}

\begin{document}
\maketitle

\begin{strip}
\vspace{-5.5em}
\begin{center}
\href{https://github.com/Yiyang-Ian-Li/LongDA}{
\raisebox{-0.12em}{\includegraphics[height=1.05em]{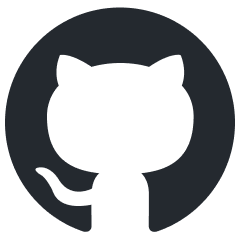}}
\;\textbf{Code}}
\qquad
\href{https://huggingface.co/datasets/EvilBench/LongDA}{
\raisebox{-0.25em}{\includegraphics[height=1.3em]{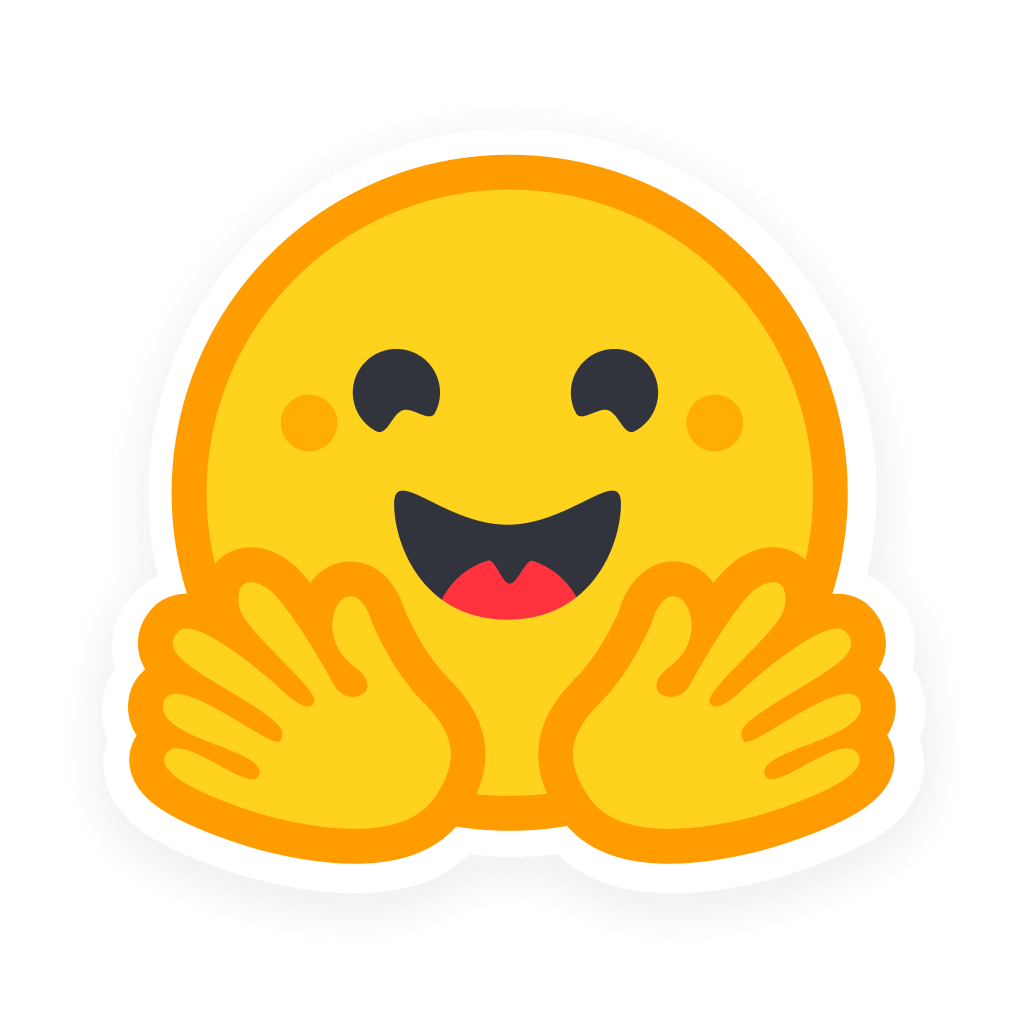}}
\;\textbf{Data}}
\end{center}
\vspace{-0.5em}
\end{strip}

\begin{abstract}
We introduce LongDA, a data analysis benchmark for evaluating LLM-based agents under documentation-intensive analytical workflows.
In contrast to existing benchmarks that assume well-specified schemas and inputs, LongDA targets real-world settings in which navigating long documentation and complex data is the primary bottleneck. 
To this end, we manually curate raw data files, long and heterogeneous documentation, and expert-written publications from 17 publicly available U.S. national surveys, from which we extract 505 analytical queries grounded in real analytical practice. 
Solving these queries requires agents to first retrieve and integrate key information from multiple unstructured documents before performing multi-step computations and writing executable code, which remains challenging for existing data analysis agents. To support the systematic evaluation under this setting, we develop LongTA, a tool-augmented agent framework that enables document access, retrieval, and code execution, and evaluate a range of proprietary and open-source models. 
Our experiments reveal substantial performance gaps even among state-of-the-art models, highlighting the challenges researchers should consider before applying LLM agents for decision support in real-world, high-stakes analytical settings.
\end{abstract}

\section{Introduction}
Recent advances in large language models (LLMs)~\cite{chang2024survey, naveed2025comprehensive} have enabled the development of LLM agents capable of assisting with data analysis tasks, including writing Python code~\cite{hollmann2023large, guo2024ds, sun2025lambda}, executing statistical computations~\cite{hong2025data}, and generating analytical reports~\cite{zhang2025deepanalyze}. These agents show strong potential to reduce the manual effort required in data-driven research and decision-making, particularly in areas such as social science, economics, and public policy, where analytical workloads commonly involve complex and large-scale datasets~\cite{sun2025survey, rahman2025llm, wang2025large}. Accordingly, benchmarking the data analysis capabilities of LLM agents has become an important and active research direction~\cite{zhang2024benchmarking, hu2025dataset}.

\begin{figure}[t]
\centering
\includegraphics[width=\linewidth]{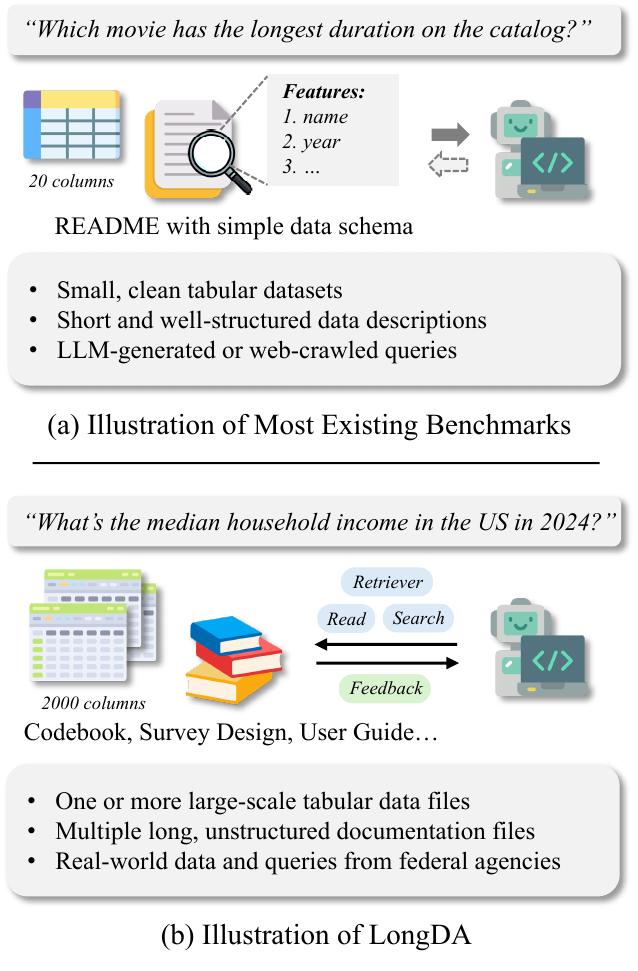}
\vspace{-15pt}
\caption{Comparison of existing data analysis benchmarks and LongDA.}
\label{fig1}
\vspace{-10pt}
\end{figure}

\begin{table*}[t]
\centering
\caption{Comparison of related benchmarks. Context denotes the approximate number of context tokens per query. Syn., Hum., and Web denote synthetic, human, and web sources, respectively. \ding{51} indicates interactive environment.}
\label{tab:benchmark_comparison}
\resizebox{\linewidth}{!}{
\begin{tabular}{l@{\hspace{15pt}} r@{\hspace{25pt}} p{2.2cm} p{2.8cm} c r}
\toprule
Benchmark & Size & Data source & Query source & Interaction & Context \\
\midrule
DS-1000~\cite{lai2023ds}                 & 1,000 & Web & Web & \ding{55} & 900 \\
Text2Analysis~\cite{he2024text2analysis} & 2,249 & Hum. & Syn. / Hum. & \ding{55} & \textsc{N/A} \\
DSCodeBench~\cite{ouyang2025dscodebench} & 1,000 & Web & Syn. / Hum. & \ding{55} & 640 \\
DataSciBench~\cite{zhang2025datascibench}& 222   & \textsc{N/A} & Syn. / Hum. / Web & \ding{55} & 200 \\
\midrule
InfiAgent-DABench~\cite{hu2024infiagent} & 257   & Web & Syn.  & \ding{51} & 160 \\
DSEval~\cite{zhang2024benchmarking}      & 825   & Web & Syn. / Hum. & \ding{51} & 1,300 \\
DACO~\cite{wu2024daco}                   & 1,942 & Web & Syn. / Hum. & \ding{51} & 890 \\
DA-Code~\cite{huang2024code}             & 500   & Web & Hum. & \ding{51} & 370 \\
DSBench~\cite{jing2024dsbench}           & 466   & Web & Web & \ding{51} & 1,060 \\
DABstep~\cite{egg2025dabstep}            & 450   & Company & Company & \ding{51} & 5,200 \\
\midrule
\textbf{LongDA (ours)}                   & \textbf{505} & \textbf{Gov.\ agencies} & \textbf{Expert pubs.} & \textbf{\ding{51}} & \textbf{263,500} \\
\bottomrule
\end{tabular}
}
\vspace{-8pt}
\end{table*}
Despite these advances, real-world data analysis remains fundamentally challenging. In practice, analytical workflows rarely begin with clean, well-documented datasets. Instead, analysts typically first navigate lengthy and heterogeneous documentation to understand dataset structures, variable definitions, and data preprocessing procedures~\cite{koesten2021talking}. This information is commonly scattered across multiple unstructured documents, such as codebooks, technical reports, and methodological appendices, which can span hundreds of thousands of tokens~\cite{ale2025enhancing}. Analysts generally need to complete this information-gathering stage before writing code to extract relevant variables and compute the results. For both humans and LLM agents, this documentation navigation stage is frequently the dominant bottleneck in real-world data analysis.

However, most existing data analysis benchmarks do not adequately capture this reality. As illustrated in Figure~\ref{fig1} and summarized in Table~\ref{tab:benchmark_comparison}, prior benchmarks primarily evaluate data analysis under well-specified inputs, implicitly assuming that relevant variables, schemas, and data files are already known. Consequently, they overlook a critical component of real-world analysis: the ability to retrieve, interpret, and integrate information from long, unstructured documentation. Moreover, many benchmarks rely on LLM-generated or web-crawled queries, which may not reflect the analytical questions practitioners actually seek to answer. A more detailed discussion of related work is provided in Appendix~\ref{app:related_work}.

Motivated by these limitations, we introduce \textbf{LongDA}, the first data analysis benchmark that evaluates LLM agents under documentation-intensive analytical workflows. LongDA is constructed from federal agency–released data files, extensive documentation with an average length of \emph{263k tokens}, and expert-written analytical publications. From these publications, we extract \emph{505 real analytical queries} whose reported results are reproducible using publicly released data. To solve these queries, agents need to identify relevant variables, assign weights, and define populations before performing multi-step computations to obtain final answers. This remains a fundamental challenge for existing data analysis agents that are not designed to operate in this specific setting.

To enable systematic and controlled study of this challenge, we further develop \textbf{LongTA}, a lightweight tool-augmented agent framework that enables document access, retrieval, and code execution, serving as an evaluation scaffold and baseline for LongDA. 
Using LongTA, we conduct extensive experiments on a range of proprietary and open-source models, revealing substantial performance gaps and underscoring the limitations of current LLM agents in real-world analytical workflows.

Our contributions are summarized as follows:
\begin{itemize}
    \item We introduce \textbf{LongDA}, a benchmark designed to evaluate data analysis agents on realistic datasets with long, complex documentation.
    \item We develop \textbf{LongTA}, a lightweight tool-augmented baseline for controlled document access and code execution for LongDA.
    \item We extensively evaluate state-of-the-art models, revealing substantial performance gaps and identifying information retrieval and tool-use strategy as primary bottlenecks.
\end{itemize}

\begin{figure*}[t!]
\centering
\includegraphics[width=\textwidth]{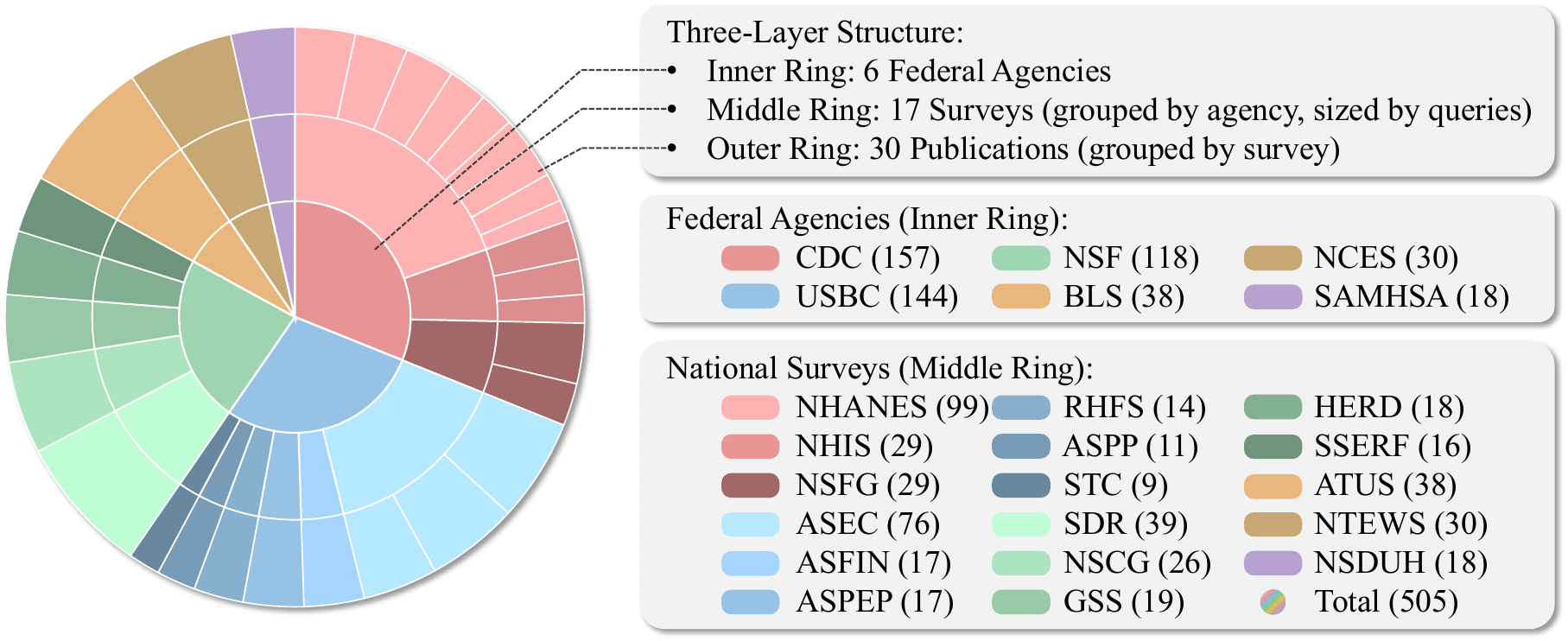}
\vspace{-10pt}
\caption{Hierarchical composition of the LongDA benchmark. Numbers in parentheses indicate the number of queries associated with each entity, with a total of 505 queries. Publications share the same color as their parent survey, and surveys belonging to the same federal agency are rendered in the same color family.}
\label{fig2}
\vspace{-10pt}
\end{figure*}

\section{LongDA Benchmark}

\subsection{Data Collection}
To construct a realistic and practically meaningful data analysis benchmark, we collect seventeen U.S.\ national surveys spanning a broad range of domains, including public health, labor and employment, education, and the scientific workforce. 
These surveys are meticulously designed, conducted, and released by six U.S. federal agencies. 
They are routinely utilized by governments, researchers, and policymakers to provide valuable insights that inform public policy, social programs, and scientific research. 
For the complete list of survey abbreviations, official names, and source links, refer to Appendix~\ref{app:survey_metadata}.

For each survey, we obtain three types of official resources from the corresponding agency websites: \textit{(1) raw data files}, \textit{(2) survey documentation}, and \textit{(3) analytical publications} produced by domain experts. The documentation typically includes codebooks, methodological descriptions, and survey design reports, which are written for human analysts and commonly span long, unstructured texts. To ensure consistency and reproducibility, we collect data, documentation, and publications from a fixed survey year for each survey and organize all resources into a unified directory structure without modifying their original content. For the complete mapping between surveys and their associated publications, refer to Appendix~\ref{app:survey_pub_mapping}.

While all three resource types are collected during dataset construction, they play different roles in the benchmark: the publications are used exclusively for query extraction and are not accessible to agents during evaluation, while raw data files and the associated documentation are exposed to the agents when solving tasks. This design ensures that benchmark queries are grounded in real analytical practice while preventing agents from directly retrieving answers from expert-written reports.

\subsection{Query and Ground Truth Extraction}
After data collection, we extract queries and ground truths from the official analytical publications.
Each query in LongDA consists of three components: \textit{(1) a question}, \textit{(2) an answer structure}, and \textit{(3) additional information}. 
The \textit{question} specifies the analytical goal based on the given survey. 
The \textit{answer structure} defines the expected format of the response, which is either a single numerical value or a fixed-length list with predefined semantic elements (e.g., [\,\textit{men}, \textit{women}\,], [\,\textit{under18}, \textit{age18\_64}, \textit{age65plus}\,]). 
The \textit{additional information} provides auxiliary context necessary for correctly interpreting the question. Together, these three components are the minimal information required to make each query both unambiguous for automatic evaluation and faithful to real-world analytical practice.

For each publication, we first verify that the reported results can be reproduced using the publicly released survey data, which is typically stated explicitly in the publication. We then construct queries based on reported numerical findings, together with their corresponding answer structures. When specifically noted, we extract relevant definitions or clarifications provided in the publication and include them as additional information to resolve potential linguistic ambiguities. This design ensures that queries are grounded in real-world analytical practice while being technically solvable, and encourages agents to perform genuine reasoning over data and documentation.

\subsection{Benchmark Statistics}
LongDA contains 505 queries in total. Figure~\ref{fig2} illustrates the hierarchical composition of LongDA, organized across three levels: federal agencies, surveys, and publications. In total, LongDA covers 6 U.S. federal agencies, 17 national surveys, and 30 official publications, reflecting the broad coverage of real-world data analysis scenarios. 

A key characteristic of LongDA is the complexity of its accompanying documentation: with an average length of 263k tokens and a maximum length exceeding 735k tokens, LongDA possesses substantially longer contexts than existing benchmarks. The distribution of documentation lengths is shown in Figure~\ref {fig:performance_across_surveys}. LongDA queries also exhibit structured answer formats: 221 queries (44\%) require a single numerical answer, while the remaining 284 queries (56\%) involve fixed-length lists with predefined semantic elements. 

\subsection{Evaluation Protocol}
To control evaluation cost and better reflect realistic analytical workflows, queries derived from the same publication are presented to the agent in \emph{blocks}, where each block contains all queries associated with a single publication (see Appendix~\ref{app:block_example}). Queries within a block typically revolve around shared themes or variables, allowing the agent to reuse intermediate understanding across related questions. Note that during evaluation, agents are restricted to accessing the raw data files and survey documentation corresponding to the given survey.

We evaluate agent performance along two primary dimensions: \emph{effectiveness} and \emph{efficiency}. 
Effectiveness is measured by \textbf{coverage rate} and \textbf{match rate}.
Efficiency is characterized by total token consumption, total runtime over the full benchmark, and the average number of interaction steps per block.
We describe each metric in detail below.

To demonstrate the agents' ability to follow task specifications and produce reliable, structured output, we compute the coverage rate, defined as the proportion of queries for which the answer syntactically conforms to the specified answer structure. Formally, given a query set $\mathcal{Q}$ and an agent’s output set $\hat{\mathcal{A}}$, the coverage rate is defined as
\[
\text{Coverage} = \frac{|\{\,q \in \mathcal{Q} \mid \hat{a}_q \text{ is structurally valid}\,\}|}{|\mathcal{Q}|},
\]
where $\hat{a}_q$ denotes the agent’s answer to query $q$.

The match rate measures numerical correctness under a predefined tolerance. 
For each query $q$ with ground truth answer $a_q$ and agent prediction $\hat{a}_q$, we define a tolerance parameter $\epsilon$ and set tolerance as
\[
\tau(a_q) = \max(\epsilon \cdot |a_q|, 1),
\]
ensuring that small-magnitude targets are not overly penalized.
The prediction is considered correct if $|\hat{a}_q - a_q| \leq \tau(a_q)$ (applied element-wise for list answers). 
The match rate is the proportion of queries for which this condition holds. 

\begin{figure*}[t!]
\centering
\includegraphics[width=\textwidth]{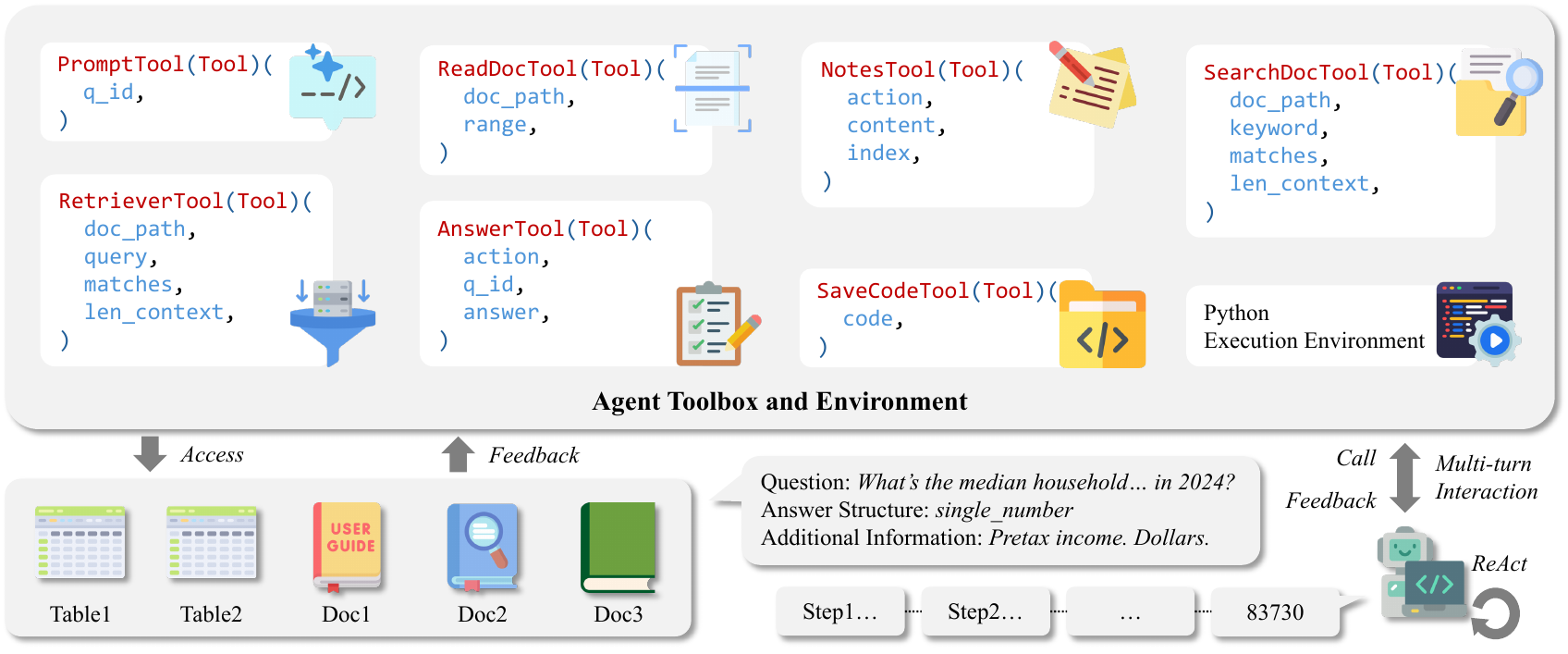}
\vspace{-10pt}
\caption{Illustration of LongTA. The agent solves queries through a ReAct-style multi-turn interaction, coordinating multiple tools for long-document navigation and code execution over heterogeneous data sources.}
\label{fig3}
\vspace{-10pt}
\end{figure*}

\section{LongTA Framework}
\label{sec:4}
To enable systematic study of LLM-based data analysis agents under documentation-intensive settings, we introduce LongTA, a lightweight tool-augmented agent framework that serves as an evaluation scaffold and baseline for LongDA. 
LongTA consists of a ReAct-style~\cite{yao2022react} coding agent and a set of specialized tools for interacting with raw data files and documentation.  

\subsection{ReAct Coding Agent}
\label{sec:react_agent}
The coding agent follows an iterative ReAct loop in which the LLM interleaves tool calls for document navigation and Python code execution for numerical computation. 
This design matches realistic analysis workflows, where an analyst repeatedly consults documentation to identify variables and then runs code to examine the statistical results.

\paragraph{Block-level prompt.}
LongDA presents queries in blocks. For each block, we construct a prompt that contains:
(i) global instructions,
(ii) a list of questions,
(iii) the expected answer structure for each question,
(iv) additional information for each question,
and (v) the list of available data files and documentation files for this survey. 
The prompt template is provided in Appendix \ref{app:prompt_template}.

\paragraph{Iterative reasoning-and-action loop.}
Given the prompt, the agent runs for at most a fixed step budget. At each step, the agent generates a Python snippet to call tools or run analysis code. The loop terminates when the agent signals task completion or when the maximum step budget is reached.

\paragraph{Execution environment and safety.}
To support common survey analysis workloads while keeping the execution controlled, we configure the executor with an allowlist of commonly-used Python libraries (e.g., \texttt{pandas}, \texttt{numpy}, \texttt{scipy}, \texttt{statsmodels}) and standard library modules (e.g., \texttt{json}, \texttt{re}), balancing practicality and containment.

\subsection{Tool Design}
\label{sec:tool_design}

LongDA tasks are bottlenecked by documentation navigation: the agent must discover variable definitions, survey weights, and file schemas from long, heterogeneous documents before writing accurate analysis code. 
Accordingly, we design a small set of tools for document interaction, context management, and answer recording, built on top of a unified internal representation of documentation.

\paragraph{Unified document representation.}
All documentation files are exposed through a unified ``document unit'' abstraction. Given a document file, we parse it into a sequence of units depending on file type:
PDFs are split into pages; text-like formats (TXT/PY/JSON) are split into lines; tabular formats (CSV/XLSX) are split into rows.
Each unit has an identifier (e.g., ``Page 3'', ``Line 120'', ``Row 15'') along with the normalized text content. This design supports both keyword search and retrieval over large files while keeping tool outputs compact.

\paragraph{Document access and retrieval.}
To navigate long, heterogeneous documentation, we provide tools that support both targeted inspection and scalable retrieval.
\textbf{(1) \texttt{read\_doc}} enables direct inspection by returning either a short preview (when no range is specified) or user-selected units via one-based ranges (e.g., \texttt{1-3} or \texttt{2,5,7}), which is useful for reading codebook sections, methodological descriptions, or specific line/page ranges.
Complementing this, \textbf{(2) \texttt{search\_doc}} performs exact keyword search across units and returns snippets with a configurable context window, allowing the agent to quickly locate candidate variable names, weighting terms, or table captions.
Since exact matching may be insufficient when the agent does not know the precise phrasing used in documentation, we further implement \textbf{(3) \texttt{retriever}}, a lightweight lexical retriever based on BM25~\cite{robertson2009probabilistic}: it splits each unit into overlapping chunks with configurable chunk size and overlap, builds index over all chunks, and returns the top-$k$ relevant chunks. 

\paragraph{Prompt revisiting and note-taking.}
To mitigate context loss in long interactions, we provide \textbf{\texttt{prompt}} for revisiting task specifications and \textbf{\texttt{notes}} for externalizing discovered information such as variable definitions and weight usage, which can be reused across queries within a block.

\begin{table*}[htbp!]
\centering
\caption{Model performance comparison. Best results for each category are highlighted in \textbf{bold}. In (M), Out (M), and Total (M) denote the numbers of input tokens, output tokens, and total tokens (in millions), respectively.}
\label{tab:main results}
\resizebox{\linewidth}{!}{
\begin{tabular}{l
                S[table-format=2.2]
                S[table-format=2.2]
                S[table-format=2.2]
                S[table-format=3.2]
                S[table-format=1.2]
                S[table-format=3.2]
                S[table-format=2.2]}
\toprule
Model & {Coverage (\%)} & {Match (\%)} & {Avg. Steps} &
{In (M)} & {Out (M)} & {Total (M)} & {Time (h)} \\
\midrule
GPT-5 (High) & \bfseries 94.65 & 68.91 & 6.20 & 7.38 & 0.87 & 8.25 & 5.74 \\
GPT-5 & 91.09 & \bfseries 69.16 & 5.50 & 5.75 & 0.65 & 6.40 & 3.58 \\
GPT-5-mini & 69.50 & 40.81 & 8.63 & 10.10 & 0.56 & 10.66 & 3.42 \\
\midrule
DeepSeek-V3.2 & \bfseries 67.33 & \bfseries 53.00 & 66.30 & 68.50 & 0.40 & 68.91 & 5.22 \\
DeepSeek-V3.2-Thinking & 61.58 & 47.82 & 37.60 & 35.55 & 0.83 & 36.38 & 9.08 \\
Kimi-K2-0905 & 50.50 & 28.27 & 36.13 & 37.78 & 0.43 & 38.21 & 5.22 \\
Qwen3-235B-A22B-Instruct & 48.91 & 27.17 & 42.30 & 61.20 & 0.53 & 61.73 & 10.38 \\
Qwen3-Coder-480B-A35B-Instruct & 51.49 & 23.44 & 53.27 & 75.13 & 0.36 & 75.49 & 5.08 \\
Qwen3-Next-80B-A3B-Instruct & 49.50 & 21.85 & 21.50 & 22.00 & 0.37 & 22.36 & 1.64 \\
GLM-4.7 & 27.13 & 19.18 & 81.17 & 119.42 & 0.59 & 120.01 & 8.42 \\
GPT-OSS-120B & 46.53 & 12.15 & 18.60 & 13.98 & 0.68 & 14.66 & 3.67 \\
\bottomrule
\end{tabular}
}
\vspace{-10pt}
\end{table*}

\paragraph{Answer recording and auditing.}
Automatic evaluation in LongDA requires answers to be structurally valid. We therefore enforce that final outputs are emitted through an \textbf{\texttt{answer}} tool. The tool validates that each answer is either a single number or a list of numbers, and stores the raw serialized value. This design separates reasoning traces from final outputs, enabling reliable automatic evaluation and post-hoc analysis.
Finally, agents can use \textbf{\texttt{save\_code}} to store the Python code used for a block, providing an audit trail that supports reproducibility checks and qualitative error analysis.

\section{Experiments}

\subsection{Experiment Settings}
We evaluate LongDA on a diverse set of state-of-the-art proprietary and open-source LLMs, including GPT-5~\cite{openai2025gpt5}, DeepSeek-V3.2~\cite{liu2025deepseek}, Qwen3 series~\cite{yang2025qwen3}, Kimi-K2~\cite{team2025kimi}, GLM-4.7~\cite{glm47_zai}, and GPT-OSS-120B~\cite{openai2025gptoss120b}. 
All models are evaluated over LongTA with a budget of 100 steps and a tolerance threshold of $5\%$ when computing the match rate.

\subsection{Main Results}

\begin{figure}[t!]
\centering
\includegraphics[width=\linewidth]{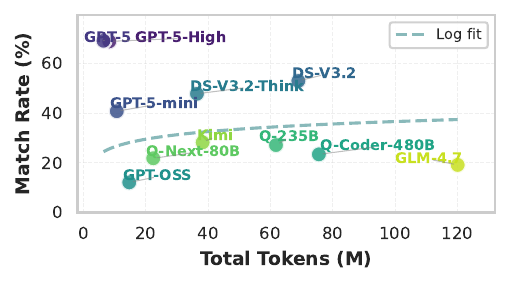}
\vspace{-20pt}
\caption{Match rate vs. total token consumption.}
\label{fig:match_rate_vs_tokens}
\vspace{-5pt}
\end{figure}

\paragraph{Overall Performance and Efficiency.}
Table~\ref{tab:main results} reveals substantial differences in both effectiveness and efficiency across models. 
The GPT-5 family achieves the highest coverage and match rates while requiring significantly fewer reasoning steps, fewer total tokens, and less wall-clock time. 
We attribute the strong performance of GPT-5 to three complementary factors.
First, GPT-5 appears to possess richer prior knowledge of survey conventions and variable nomenclature, which often allows it to rapidly infer or approximate variable abbreviations before consulting documentation. 
This behavior is also reflected in our case study (Appendix~\ref{app:case_study}), where GPT-5 identifies many relevant keywords within the first few steps.
Second, GPT-5 demonstrates substantially stronger capability in strategically leveraging the \texttt{search\_doc} tool for information retrieval, as further analyzed in Section~\ref{sec:tool_usage}. 
Third, GPT-5 tends to execute larger code blocks per step (combining multiple keyword searches and sub-tasks within a single execution) rather than issuing one fine-grained tool call step by step. 
This execution style increases the effective information density of the context and avoids repeatedly injecting system prompts and intermediate states, thereby achieving both higher efficiency and stronger performance.
Notably, even GPT-5 (High) achieves only $68.91\%$ match rate, indicating substantial remaining headroom. 

Among open-source models, DeepSeek-V3.2 exhibits comparatively stable generation behavior. However, we also observe that other models (e.g., Kimi and GLM series) sometimes fall into repetitive generation patterns. 
Such repetition substantially inflates output token usage and runtime, contributing to lower overall efficiency.

\begin{figure}[t!]
\centering
\includegraphics[width=\linewidth]{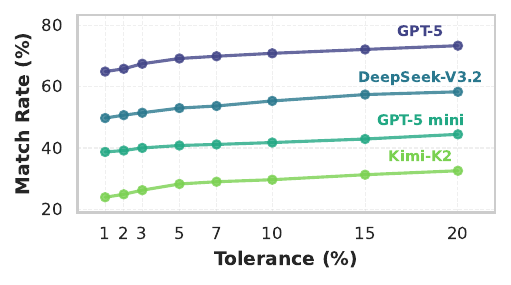}
\vspace{-20pt}
\caption{Match rate sensitivity to tolerance threshold.}
\label{fig:match_rate_vs_tolerance}
\vspace{-5pt}
\end{figure}

\begin{figure*}[t!]
\centering
\includegraphics[width=\linewidth]{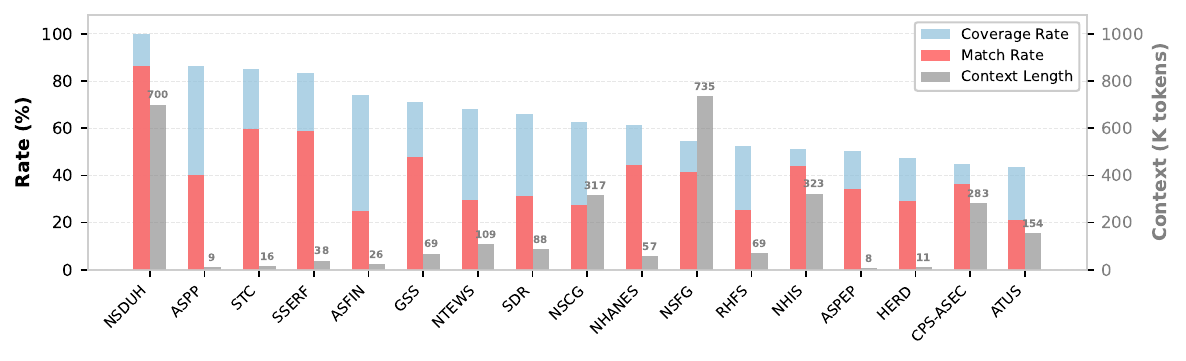}
\vspace{-25pt}
\caption{Survey-level performance averaged across all evaluated models.}
\label{fig:performance_across_surveys}
\vspace{-10pt}
\end{figure*}

\begin{figure}[htbp]
\centering
\includegraphics[width=\linewidth]{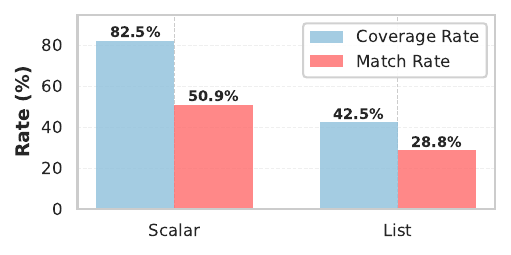}
\vspace{-20pt}
\caption{Performance by answer type (scalar vs.\ list).}
\vspace{-10pt}
\label{fig:scalar_vs_list}
\end{figure}

\paragraph{Scaling Behavior.}
Figure~\ref{fig:match_rate_vs_tokens} illustrates the relationship between match rate and total token consumption. 
To analyze scaling behavior, we fit a logarithmic curve to the data. Note that GPT-5 and GPT-5 (High) are excluded from the fit as outliers to avoid distortion of the overall trend.
Performance improves with increased token budget, though the gains exhibit diminishing returns, reflecting the intrinsic difficulty of LongDA.

\paragraph{Tolerance Sensitivity.}
Figure~\ref{fig:match_rate_vs_tolerance} studies sensitivity to the tolerance threshold. 
While match rate increases monotonically with tolerance, the relative ranking of models and the overall performance gaps remain stable, indicating that the evaluation is not overly sensitive to the exact tolerance choice.

\paragraph{Survey-Level Analysis.}
Figure~\ref{fig:performance_across_surveys} reports survey-level performance aggregated across all evaluated models, showing a mild negative correlation between documentation length and model performance.
NSDUH is a notable exception: despite extremely long contexts, it achieves comparatively strong performance. 
This is likely because its publicly available microdata already includes many pre-coded variables derived from multiple raw fields, allowing models to answer queries once relevant variables are identified without requiring complex downstream computation.

\paragraph{Scalar vs.\ List Queries.}
Figure~\ref{fig:scalar_vs_list} compares scalar and list-structured queries. 
List queries are consistently more difficult than scalar ones, due to both increased computational complexity and the requirement to follow specific answer structures.

\paragraph{Effect of Explicit Reasoning.}
Comparing GPT-5 and DeepSeek-V3.2 with and without explicit reasoning variants, we observe that explicit reasoning does not provide clear performance improvements on LongDA.
This suggests that the core difficulty of LongDA lies not in logical inference, but in discovering and extracting sufficient information from long, complex documentation.
Moreover, reasoning-oriented models exhibit higher tool usage, longer outputs, and increased runtime (Table~\ref{tab:main results} and~\ref{tab:tool_count}). We hypothesize that cautious reasoning induces excessive tool calls without substantially improving document understanding, introducing overhead with limited accuracy gains.

\begin{figure*}[t!]
\centering
\includegraphics[width=\linewidth]{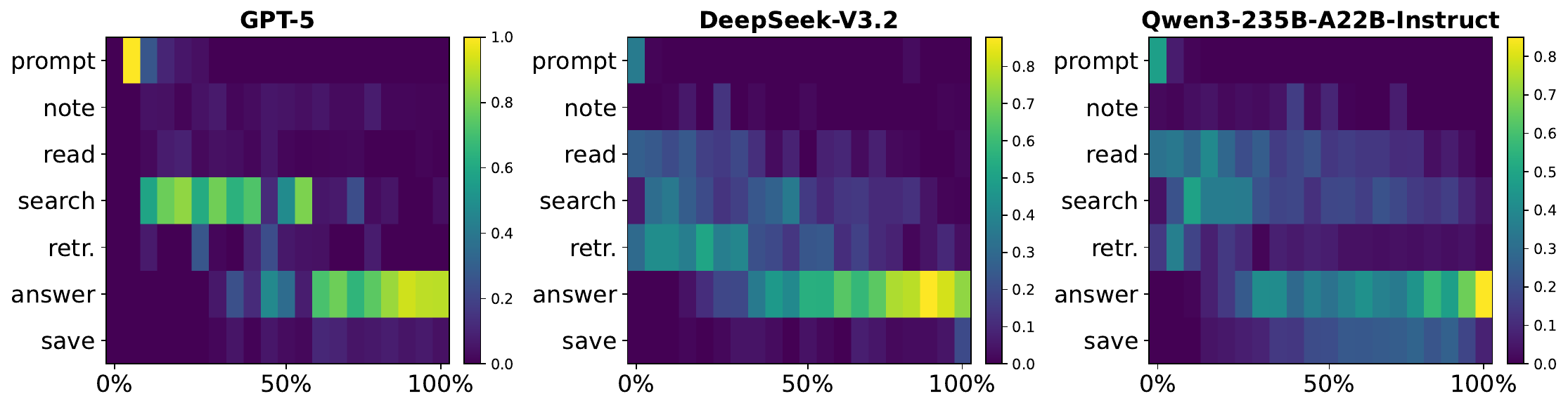}
\vspace{-20pt}
\caption{Normalized tool usage trajectories across three agent models. Each heatmap shows the probability distribution of tool calls over the normalized problem-solving progress (0–100\%).}
\label{fig:tool_trajectory}
\end{figure*}

\begin{table*}[htbp!]
\centering
\caption{Tool usage count by models.}
\label{tab:tool_count}
\resizebox{0.85\linewidth}{!}{
\begin{tabular}{l
                S[table-format=2.0]
                S[table-format=3.0]
                S[table-format=4.0]
                S[table-format=3.0]
                S[table-format=2.0]
                S[table-format=3.0]
                S[table-format=3.0]}
\toprule
Model & {Prompt} & {Read} & {Search} & {Retr.} & {Note} & {Answer} & {Save} \\
\midrule
GPT-5 & 30 & 46 & 1704 & 114 & 79 & 704 & 79 \\
GPT-5 (High) & 30 & 121 & 2392 & 117 & 77 & 754 & 74 \\
DeepSeek-V3.2 & 31 & 146 & 207 & 269 & 38 & 526 & 39 \\
DeepSeek-V3.2-Thinking & 30 & 122 & 206 & 393 & 39 & 893 & 95 \\
Qwen3-235B-A22B-Instruct & 30 & 309 & 371 & 146 & 57 & 596 & 293 \\
Qwen3-Coder-480B-A35B-Instruct & 84 & 187 & 231 & 15 & 17 & 528 & 18 \\
\bottomrule
\end{tabular}
}
\end{table*}
\begin{table*}[t!]
\centering
\caption{Ablation study on tool availability.}
\label{tab:ablation_no_retriever}
\resizebox{\linewidth}{!}{
\begin{tabular}{l l
                S[table-format=2.2]
                S[table-format=2.2]
                S[table-format=2.2]
                S[table-format=3.2]
                S[table-format=1.2]
                S[table-format=2.2]}
\toprule
Model & Setting & {Coverage (\%)} & {Match (\%)} & {Steps} &
{In (M)} & {Out (M)} & {Time (h)} \\
\midrule
GPT-5 & Full tools & \bfseries 91.09 & \bfseries 69.16 & 5.50 & 5.75 & 0.65 & 3.58 \\
GPT-5 & w/o \texttt{search\_doc} & 80.20 & 58.30 & 6.03 & 6.09 & 0.65 & 2.84 \\
\midrule
DeepSeek-V3.2 & Full tools & \bfseries 67.33 & 53.00 & 66.30 & 68.50 & 0.40 & 5.22 \\
DeepSeek-V3.2 & w/o \texttt{retriever} & 66.93 & \bfseries 54.03 & 65.07 & 67.04 & 0.40 & 5.36 \\
\midrule
Qwen3-Coder-480B-A35B-Instruct & Full tools & 51.49 & \bfseries 23.44 & 53.27 & 75.13 & 0.36 & 5.08 \\
Qwen3-Coder-480B-A35B-Instruct & w/o \texttt{retriever} & \bfseries 53.27 & 21.52 & 48.00 & 79.48 & 0.40 & 4.27 \\
\bottomrule
\end{tabular}
}
\end{table*}

\subsection{Tool Usage Analysis}
\label{sec:tool_usage}

Table~\ref{tab:tool_count} reports aggregate tool usage statistics. 
Additionally, to visualize tool-use dynamics, we extract full tool-call traces from agent execution logs and represent each block's trajectory as a sequence of tool invocations. 
Each trajectory is normalized to a $[0,1]$ progress scale based on its maximum step count and discretized into 20 bins. 
For each bin, we compute the empirical distribution of tool usage across all trajectories of a model, producing the heatmaps in Figure~\ref{fig:tool_trajectory}.

\paragraph{Behavioral Analysis.}
GPT-5 variants strongly favor the \texttt{search} tool over the \texttt{retriever} and consistently invoke the \texttt{prompt} tool at the beginning of each block.
Open-source models display more exploratory tool usage patterns with heavier reliance on retrieval and iterative refinement.
These results indicate that successful performance on LongDA depends not only on language modeling and coding ability, but also on effective tool-use strategies for navigating long, complex documentation.

We further conduct a tool ablation study, as reported in Table~\ref{tab:ablation_no_retriever}.
According to the results, removing \texttt{search\_doc} leads to a substantial degradation for GPT-5, indicating that GPT-5 is particularly effective at leveraging \texttt{search\_doc} for agentic problem solving in LongDA.
In contrast, removing \texttt{retriever} causes only marginal fluctuations for DeepSeek-V3.2 and Qwen3-Coder.
This suggests that the performance gap between GPT-5 variants and other models cannot be explained solely by a preference for the \texttt{search\_doc} tool.
Instead, LongDA primarily challenges models’ ability to strategically extract relevant evidence from long documentation and integrate retrieved information into coherent multi-step analytical workflows.

\subsection{Summary of Findings}

Overall, our experiments lead to three key observations. \textbf{First}, current state-of-the-art models remain far from solving documentation-intensive data analysis reliably. 
Even the strongest model, GPT-5 (High), achieves only 68.91\% match rate, indicating substantial headroom.
\textbf{Second}, success on LongDA is driven primarily by information retrieval and tool-use strategy rather than by pure reasoning capability.
Models that are better at retrieving relevant evidence from long documentation and orchestrating tool calls consistently outperform others.
\textbf{Third}, longer contexts and more complex answer structures amplify difficulty, reflecting the challenges of information retrieval, instruction following, and multi-step computation.

\section{Conclusion}
We introduce LongDA, the first benchmark for evaluating data analysis agents under documentation-intensive analytical workflows.  
By grounding tasks in large-scale government datasets and long, unstructured documentation, LongDA exposes fundamental challenges to current LLM-based agents, particularly in information retrieval and document navigation. 
Our extensive experiments show that even state-of-the-art models struggle with documentation-intensive analytical pipelines, where performance depends critically on effective evidence retrieval and strategic tool use under long-context constraints. 
We hope LongDA will serve as a challenging and practical testbed for advancing the development of reliable data analysis agents.

\section{Limitations}
\label{sec:limitations}

LongDA has several limitations that should be considered when interpreting results and when using the benchmark.

\paragraph{Scope and representativeness.}
LongDA is constructed from 17 U.S.\ national surveys and associated publications, which provide breadth across agencies and domains but still represent a small slice of the full landscape of real-world data analysis. The benchmark may under-represent domains with substantially different data modalities (e.g., proprietary enterprise data) or workflows (e.g., streaming analytics, database-centric ETL).

\paragraph{Fixed-year snapshots and documentation drift.}
Each survey is collected from a fixed release year and organized into a static local directory. In practice, survey documentation and variable definitions can change across cycles, and agency guidelines may be updated. LongDA does not explicitly test robustness to cross-year schema drift or versioned documentation.

\paragraph{Evaluation focuses on final numbers.}
Our primary metrics (coverage and match) evaluate whether agents produce structurally valid outputs and numerically correct answers within a tolerance. This design supports scalable automatic evaluation, but it does not directly measure intermediate analytical quality (e.g., whether the agent’s code follows best practices, produces confidence intervals, or correctly handles edge cases) nor does it capture the quality of explanations and reporting narratives.

\paragraph{Survey design complexity is only partially exercised.}
While LongDA tasks require reading documentation and applying sample weights, many survey analyses in practice also involve variance estimation and design-based uncertainty quantification (e.g., strata/PSU adjustments, replicate weights, finite population corrections, and survey-specific variance estimators). LongDA currently does not include queries that explicitly evaluate variance estimation (e.g., standard errors or confidence intervals). As a result, strong performance on LongDA does not imply full competence in general complex-survey inference beyond point estimation.

\section{Ethical Considerations}
LongDA is constructed from publicly released U.S.\ government survey data and documentation. 
The benchmark does not include personally identifiable information (PII) and does not require any individual-level inference. 
Given the high-stakes nature of real-world data analysis, outputs produced by LLM agents should not be used for automated decision-making without human verification. 
We report token usage and runtime to make evaluation cost explicit and to encourage more efficient and reproducible methods.

\bibliography{main}

@inproceedings{lai2023ds,
  title={DS-1000: A natural and reliable benchmark for data science code generation},
  author={Lai, Yuhang and Li, Chengxi and Wang, Yiming and Zhang, Tianyi and Zhong, Ruiqi and Zettlemoyer, Luke and Yih, Wen-tau and Fried, Daniel and Wang, Sida and Yu, Tao},
  booktitle={International Conference on Machine Learning},
  pages={18319--18345},
  year={2023},
  organization={PMLR}
}

@article{hu2024infiagent,
  title={Infiagent-dabench: Evaluating agents on data analysis tasks},
  author={Hu, Xueyu and Zhao, Ziyu and Wei, Shuang and Chai, Ziwei and Ma, Qianli and Wang, Guoyin and Wang, Xuwu and Su, Jing and Xu, Jingjing and Zhu, Ming and others},
  journal={arXiv preprint arXiv:2401.05507},
  year={2024}
}

@article{jing2024dsbench,
  title={DSBench: How Far Are Data Science Agents from Becoming Data Science Experts?},
  author={Jing, Liqiang and Huang, Zhehui and Wang, Xiaoyang and Yao, Wenlin and Yu, Wenhao and Ma, Kaixin and Zhang, Hongming and Du, Xinya and Yu, Dong},
  journal={arXiv preprint arXiv:2409.07703},
  year={2024}
}

@article{egg2025dabstep,
  title={DABstep: Data Agent Benchmark for Multi-step Reasoning},
  author={Egg, Alex and Goyanes, Martin Iglesias and Kingma, Friso and Mora, Andreu and von Werra, Leandro and Wolf, Thomas},
  journal={arXiv preprint arXiv:2506.23719},
  year={2025}
}

@article{huang2024code,
  title={Da-code: Agent data science code generation benchmark for large language models},
  author={Huang, Yiming and Luo, Jianwen and Yu, Yan and Zhang, Yitong and Lei, Fangyu and Wei, Yifan and He, Shizhu and Huang, Lifu and Liu, Xiao and Zhao, Jun and others},
  journal={arXiv preprint arXiv:2410.07331},
  year={2024}
}

@article{wu2024daco,
  title={DACO: Towards Application-Driven and Comprehensive Data Analysis via Code Generation},
  author={Wu, Xueqing and Zheng, Rui and Sha, Jingzhen and Wu, Te-Lin and Zhou, Hanyu and Tang, Mohan and Chang, Kai-Wei and Peng, Nanyun and Huang, Haoran},
  journal={Advances in Neural Information Processing Systems},
  volume={37},
  pages={90661--90682},
  year={2024}
}

@article{ouyang2025dscodebench,
  title={DSCodeBench: A Realistic Benchmark for Data Science Code Generation},
  author={Ouyang, Shuyin and Huang, Dong and Guo, Jingwen and Sun, Zeyu and Zhu, Qihao and Zhang, Jie M},
  journal={arXiv preprint arXiv:2505.15621},
  year={2025}
}

@article{zhang2025datascibench,
  title={Datascibench: An llm agent benchmark for data science},
  author={Zhang, Dan and Zhoubian, Sining and Cai, Min and Li, Fengzu and Yang, Lekang and Wang, Wei and Dong, Tianjiao and Hu, Ziniu and Tang, Jie and Yue, Yisong},
  journal={arXiv preprint arXiv:2502.13897},
  year={2025}
}

@inproceedings{zhang2024benchmarking,
  title={Benchmarking data science agents},
  author={Zhang, Yuge and Jiang, Qiyang and XingyuHan, XingyuHan and Chen, Nan and Yang, Yuqing and Ren, Kan},
  booktitle={Proceedings of the 62nd Annual Meeting of the Association for Computational Linguistics (Volume 1: Long Papers)},
  pages={5677--5700},
  year={2024}
}

@inproceedings{he2024text2analysis,
  title={Text2analysis: A benchmark of table question answering with advanced data analysis and unclear queries},
  author={He, Xinyi and Zhou, Mengyu and Xu, Xinrun and Ma, Xiaojun and Ding, Rui and Du, Lun and Gao, Yan and Jia, Ran and Chen, Xu and Han, Shi and others},
  booktitle={Proceedings of the AAAI Conference on Artificial Intelligence},
  volume={38},
  number={16},
  pages={18206--18215},
  year={2024}
}

@article{zhang2025deepanalyze,
  title={Deepanalyze: Agentic large language models for autonomous data science},
  author={Zhang, Shaolei and Fan, Ju and Fan, Meihao and Li, Guoliang and Du, Xiaoyong},
  journal={arXiv preprint arXiv:2510.16872},
  year={2025}
}

@article{hollmann2023large,
  title={Large language models for automated data science: Introducing caafe for context-aware automated feature engineering},
  author={Hollmann, Noah and M{\"u}ller, Samuel and Hutter, Frank},
  journal={Advances in Neural Information Processing Systems},
  volume={36},
  pages={44753--44775},
  year={2023}
}

@article{guo2024ds,
  title={Ds-agent: Automated data science by empowering large language models with case-based reasoning},
  author={Guo, Siyuan and Deng, Cheng and Wen, Ying and Chen, Hechang and Chang, Yi and Wang, Jun},
  journal={arXiv preprint arXiv:2402.17453},
  year={2024}
}

@article{sun2025lambda,
  title={Lambda: A large model based data agent},
  author={Sun, Maojun and Han, Ruijian and Jiang, Binyan and Qi, Houduo and Sun, Defeng and Yuan, Yancheng and Huang, Jian},
  journal={Journal of the American Statistical Association},
  pages={1--13},
  year={2025},
  publisher={Taylor \& Francis}
}

@inproceedings{hong2025data,
  title={Data interpreter: An llm agent for data science},
  author={Hong, Sirui and Lin, Yizhang and Liu, Bang and Liu, Bangbang and Wu, Binhao and Zhang, Ceyao and Li, Danyang and Chen, Jiaqi and Zhang, Jiayi and Wang, Jinlin and others},
  booktitle={Findings of the Association for Computational Linguistics: ACL 2025},
  pages={19796--19821},
  year={2025}
}

@inproceedings{yao2022react,
  title={React: Synergizing reasoning and acting in language models},
  author={Yao, Shunyu and Zhao, Jeffrey and Yu, Dian and Du, Nan and Shafran, Izhak and Narasimhan, Karthik R and Cao, Yuan},
  booktitle={The eleventh international conference on learning representations},
  year={2022}
}

@article{naveed2025comprehensive,
  title={A comprehensive overview of large language models},
  author={Naveed, Humza and Khan, Asad Ullah and Qiu, Shi and Saqib, Muhammad and Anwar, Saeed and Usman, Muhammad and Akhtar, Naveed and Barnes, Nick and Mian, Ajmal},
  journal={ACM Transactions on Intelligent Systems and Technology},
  volume={16},
  number={5},
  pages={1--72},
  year={2025},
  publisher={ACM New York, NY}
}

@article{chang2024survey,
  title={A survey on evaluation of large language models},
  author={Chang, Yupeng and Wang, Xu and Wang, Jindong and Wu, Yuan and Yang, Linyi and Zhu, Kaijie and Chen, Hao and Yi, Xiaoyuan and Wang, Cunxiang and Wang, Yidong and others},
  journal={ACM transactions on intelligent systems and technology},
  volume={15},
  number={3},
  pages={1--45},
  year={2024},
  publisher={ACM New York, NY}
}

@article{rahman2025llm,
  title={Llm-based data science agents: A survey of capabilities, challenges, and future directions},
  author={Rahman, Mizanur and Bhuiyan, Amran and Islam, Mohammed Saidul and Laskar, Md Tahmid Rahman and Mahbub, Ridwan and Masry, Ahmed and Joty, Shafiq and Hoque, Enamul},
  journal={arXiv preprint arXiv:2510.04023},
  year={2025}
}

@article{sun2025survey,
  title={A survey on large language model-based agents for statistics and data science},
  author={Sun, Maojun and Han, Ruijian and Jiang, Binyan and Qi, Houduo and Sun, Defeng and Yuan, Yancheng and Huang, Jian},
  journal={The American Statistician},
  pages={1--14},
  year={2025},
  publisher={Taylor \& Francis}
}

@article{ale2025enhancing,
  title={Enhancing statistical analysis of real world data},
  author={Ale, Laha and Gentleman, Robert and Endres, Christopher and Pullman, Sam and Palmer, Nathan and Goncalves, Rafael and Sarkar, Deepayan},
  journal={Database},
  volume={2025},
  pages={baaf073},
  year={2025},
  publisher={Oxford University Press}
}

@article{koesten2021talking,
  title={Talking datasets--understanding data sensemaking behaviours},
  author={Koesten, Laura and Gregory, Kathleen and Groth, Paul and Simperl, Elena},
  journal={International journal of human-computer studies},
  volume={146},
  pages={102562},
  year={2021},
  publisher={Elsevier}
}

@article{team2025kimi,
  title={Kimi k2: Open agentic intelligence},
  author={Team, Kimi and Bai, Yifan and Bao, Yiping and Chen, Guanduo and Chen, Jiahao and Chen, Ningxin and Chen, Ruijue and Chen, Yanru and Chen, Yuankun and Chen, Yutian and others},
  journal={arXiv preprint arXiv:2507.20534},
  year={2025}
}

@article{liu2025deepseek,
  title={Deepseek-v3. 2: Pushing the frontier of open large language models},
  author={Liu, Aixin and Mei, Aoxue and Lin, Bangcai and Xue, Bing and Wang, Bingxuan and Xu, Bingzheng and Wu, Bochao and Zhang, Bowei and Lin, Chaofan and Dong, Chen and others},
  journal={arXiv preprint arXiv:2512.02556},
  year={2025}
}

@article{yang2025qwen3,
  title={Qwen3 technical report},
  author={Yang, An and Li, Anfeng and Yang, Baosong and Zhang, Beichen and Hui, Binyuan and Zheng, Bo and Yu, Bowen and Gao, Chang and Huang, Chengen and Lv, Chenxu and others},
  journal={arXiv preprint arXiv:2505.09388},
  year={2025}
}

@techreport{glm47_zai,
  title        = {GLM-4.7: Advancing the Coding Capability},
  author       = {{Z.ai}},
  institution  = {Zhipu AI},
  year         = {2025},
  url          = {https://z.ai/blog/glm-4.7},
  organization = {Z.ai}
}

@misc{hu2025dataset,
  title={A Dataset-Centric Survey of LLM-Agents for Data Science},
  author={Hu, Chuxuan and Dalal, Dwip and Zhou, Xiaona},
  year={2025},
  publisher={OpenReview}
}

@article{wang2025large,
  title={Large language model-based data science agent: A survey},
  author={Wang, Peiran and Yu, Yaoning and Chen, Ke and Zhan, Xianyang and Wang, Haohan},
  journal={arXiv preprint arXiv:2508.02744},
  year={2025}
}

@inproceedings{gao2023pal,
  title={Pal: Program-aided language models},
  author={Gao, Luyu and Madaan, Aman and Zhou, Shuyan and Alon, Uri and Liu, Pengfei and Yang, Yiming and Callan, Jamie and Neubig, Graham},
  booktitle={International Conference on Machine Learning},
  pages={10764--10799},
  year={2023},
  organization={PMLR}
}

@article{zhu2024large,
  title={Are large language models good statisticians?},
  author={Zhu, Yizhang and Du, Shiyin and Li, Boyan and Luo, Yuyu and Tang, Nan},
  journal={Advances in Neural Information Processing Systems},
  volume={37},
  pages={62697--62731},
  year={2024}
}

@inproceedings{zhang2024mlcopilot,
  title={Mlcopilot: Unleashing the power of large language models in solving machine learning tasks},
  author={Zhang, Lei and Zhang, Yuge and Ren, Kan and Li, Dongsheng and Yang, Yuqing},
  booktitle={Proceedings of the 18th Conference of the European Chapter of the Association for Computational Linguistics (Volume 1: Long Papers)},
  pages={2931--2959},
  year={2024}
}

@article{zhang2023automl,
  title={Automl-gpt: Automatic machine learning with gpt},
  author={Zhang, Shujian and Gong, Chengyue and Wu, Lemeng and Liu, Xingchao and Zhou, Mingyuan},
  journal={arXiv preprint arXiv:2305.02499},
  year={2023}
}

@article{jiang2025aide,
  title={Aide: Ai-driven exploration in the space of code},
  author={Jiang, Zhengyao and Schmidt, Dominik and Srikanth, Dhruv and Xu, Dixing and Kaplan, Ian and Jacenko, Deniss and Wu, Yuxiang},
  journal={arXiv preprint arXiv:2502.13138},
  year={2025}
}

@article{li2024autokaggle,
  title={Autokaggle: A multi-agent framework for autonomous data science competitions},
  author={Li, Ziming and Zang, Qianbo and Ma, David and Guo, Jiawei and Zheng, Tuney and Liu, Minghao and Niu, Xinyao and Wang, Yue and Yang, Jian and Liu, Jiaheng and others},
  journal={arXiv preprint arXiv:2410.20424},
  year={2024}
}

@inproceedings{wang2024executable,
  title={Executable code actions elicit better llm agents},
  author={Wang, Xingyao and Chen, Yangyi and Yuan, Lifan and Zhang, Yizhe and Li, Yunzhu and Peng, Hao and Ji, Heng},
  booktitle={Forty-first International Conference on Machine Learning},
  year={2024}
}

@article{trirat2024automl,
  title={Automl-agent: A multi-agent llm framework for full-pipeline automl},
  author={Trirat, Patara and Jeong, Wonyong and Hwang, Sung Ju},
  journal={arXiv preprint arXiv:2410.02958},
  year={2024}
}

@article{robertson2009probabilistic,
  title={The probabilistic relevance framework: BM25 and beyond},
  author={Robertson, Stephen and Zaragoza, Hugo and others},
  journal={Foundations and Trends{\textregistered} in Information Retrieval},
  volume={3},
  number={4},
  pages={333--389},
  year={2009},
  publisher={Now Publishers, Inc.}
}

@techreport{openai2025gpt5,
  title        = {GPT-5 System Card},
  author       = {{OpenAI}},
  institution  = {OpenAI},
  year         = {2025},
  month        = aug,
  url          = {https://cdn.openai.com/gpt-5-system-card.pdf}
}

@misc{openai2025gptoss120b,
  title        = {Introducing GPT-OSS and GPT-OSS-120B},
  author       = {{OpenAI}},
  year         = {2025},
  url          = {https://openai.com/index/introducing-gpt-oss/}
}

\appendix

\section{Related Work}
\label{app:related_work}

\subsection{Data Analysis Benchmarks}
Existing data analysis benchmarks can be broadly categorized into single-turn and interactive settings. Single-turn benchmarks such as DS-1000~\cite{lai2023ds}, DSCodeBench~\cite{ouyang2025dscodebench}, and DataSciBench~\cite{zhang2025datascibench} primarily evaluate one-shot code generation or analytical reasoning with well-specified prompts and structured inputs, placing limited demands on document-level language understanding.

More recent interactive benchmarks (e.g., InfiAgent-DABench~\cite{hu2024infiagent}, DSEval~\cite{zhang2024benchmarking}, DA-Code~\cite{huang2024code}, DSBench~\cite{jing2024dsbench}) allow agents to perform multi-step reasoning and interact with execution environments. However, most of these benchmarks are still built on relatively small datasets with limited or simplified documentation, and therefore do not fully capture the challenge of navigating large-scale, unstructured data descriptions. In contrast, LongDA evaluates data analysis agents on real-world datasets with extensive and complex documentation, requiring language understanding, information retrieval, and tool-assisted computation.

\subsection{LLM Data Analysis Agents}

Recent work has increasingly explored using large language models as autonomous agents for data analysis.
Early efforts primarily focused on leveraging LLMs for code generation and computational reasoning, simplifying complex analytical workflows through program synthesis and tool execution~\cite{gao2023pal, zhu2024large, wang2024executable}. Building on this foundation, a new generation of data science agents has emerged, integrating function-calling, code interpreters, and external tools to support end-to-end analytical processes, including Data Interpreter~\cite{hong2025data}, DS-Agent~\cite{guo2024ds}, LAMBDA~\cite{sun2025lambda}, and related systems~\cite{zhang2023automl, zhang2024mlcopilot, li2024autokaggle, trirat2024automl, jiang2025aide}.

Collectively, these systems demonstrate that LLM agents can effectively automate many components of data science pipelines, such as feature engineering, hyper-parameter tuning, model training, and evaluation. However, most existing approaches assume relatively clean, well-documented inputs and place limited emphasis on navigating large-scale, heterogeneous documentation. As a result, they under-exercise the document understanding and information discovery capabilities that dominate real-world analytical workflows. Moreover, scalability, long-context management, and robust multi-step planning remain open challenges for current agent frameworks.
\section{Survey Metadata}
\begin{table*}[htbp]
\centering
\caption{Survey Abbreviations, Official Names, and Data Sources}
\label{tab:survey_list}
\resizebox{\linewidth}{!}{
\begin{tabular}{llllc}
\toprule
Abbrev. & Official Survey Name & Agency & Year & Data Source \\
\midrule
ATUS & American Time Use Survey & Bureau of Labor Statistics & 2024 & \href{https://www.bls.gov/tus}{Link} \\
\midrule
NHANES & National Health and Nutrition Examination Survey & National Center for Health Statistics & 2021--2023 & \href{https://www.cdc.gov/nchs/nhanes}{Link} \\
NHIS & National Health Interview Survey & National Center for Health Statistics & 2023 & \href{https://www.cdc.gov/nchs/nhis}{Link} \\
NSFG & National Survey of Family Growth & National Center for Health Statistics & 2022--2023 & \href{https://www.cdc.gov/nchs/nsfg}{Link} \\
\midrule
GSS & Graduate Student Survey & National Science Foundation & 2023 & \href{https://ncses.nsf.gov/surveys/graduate-students-postdoctorates-s-e}{Link} \\
HERD & Higher Education R\&D Survey & National Science Foundation & 2023 & \href{https://ncses.nsf.gov/surveys/higher-education-research-development}{Link} \\
NSCG & National Survey of College Graduates & National Science Foundation & 2023 & \href{https://ncses.nsf.gov/surveys/national-survey-of-college-graduates}{Link} \\
SDR & Survey of Doctorate Recipients & National Science Foundation & 2023 & \href{https://ncses.nsf.gov/surveys/doctorate-recipients}{Link} \\
SSERF & Survey of Science \& Engineering Research Facilities & National Science Foundation & 2023 & \href{https://ncses.nsf.gov/surveys/science-engineering-research-facilities}{Link} \\
\midrule
NTEWS & National Training, Education, and Workforce Survey & National Center for Education Statistics & 2024 & \href{https://nces.ed.gov/surveys/ntews}{Link} \\
\midrule
NSDUH & National Survey on Drug Use and Health & SAMHSA & 2023 & \href{https://www.samhsa.gov/data/data-we-collect/nsduh}{Link} \\
\midrule
ASFIN & Annual Survey of State Government Finances & U.S. Census Bureau & 2021 & \href{https://www.census.gov/programs-surveys/state.html}{Link} \\
ASPEP & Annual Survey of Public Employment \& Payroll & U.S. Census Bureau & 2024 & \href{https://www.census.gov/programs-surveys/apes.html}{Link} \\
ASPP & Annual Survey of Public Pensions & U.S. Census Bureau & 2019 & \href{https://www.census.gov/programs-surveys/aspp.html}{Link} \\
CPS--ASEC & Current Population Survey -- ASEC & U.S. Census Bureau \& BLS & 2024 & \href{https://www.census.gov/programs-surveys/cps.html}{Link} \\
RHFS & Rental Housing Finance Survey & U.S. Census Bureau & 2023 & \href{https://www.census.gov/programs-surveys/rhfs.html}{Link} \\
STC & Survey of State Tax Collections & U.S. Census Bureau & 2020 & \href{https://www.census.gov/programs-surveys/stc.html}{Link} \\
\bottomrule
\end{tabular}
}
\end{table*}
\begin{table*}[htbp]
\centering
\caption{Mapping Between Surveys and Source Publications}
\label{tab:survey_pub}
\resizebox{\linewidth}{!}{
\begin{tabular}{llll}
\toprule
Survey & Publication & Agency & Year \\
\midrule
ATUS & American Time Use Survey: 2024 Results & Bureau of Labor Statistics & 2024 \\
\midrule
NHANES & Data Brief No.~508: Obesity Prevalence & National Center for Health Statistics & 2021--2023 \\
NHANES & Data Brief No.~511: Hypertension Prevalence & National Center for Health Statistics & 2021--2023 \\
NHANES & Data Brief No.~515: Sleep Duration & National Center for Health Statistics & 2021--2023 \\
NHANES & Data Brief No.~516: Physical Activity & National Center for Health Statistics & 2021--2023 \\
NHANES & Data Brief No.~519: Anemia Prevalence & National Center for Health Statistics & 2021--2023 \\
NHANES & Data Brief No.~527: Depression Prevalence & National Center for Health Statistics & 2021--2023 \\
NHANES & Data Brief No.~533: Fast Food Consumption & National Center for Health Statistics & 2021--2023 \\
NHANES & Data Brief No.~538: Prescription Opioid Use & National Center for Health Statistics & 2021--2023 \\
NHANES & Data Brief No.~540: Diabetes Prevalence & National Center for Health Statistics & 2021--2023 \\
\midrule
NHIS & Data Brief No.~518: Chronic Pain & National Center for Health Statistics & 2023 \\
NHIS & Data Brief No.~528: Depression Medication & National Center for Health Statistics & 2023 \\
NHIS & Data Brief No.~529: COPD Prevalence & National Center for Health Statistics & 2023 \\
\midrule
NSFG & Data Brief No.~520: Contraceptive Use & National Center for Health Statistics & 2022--2023 \\
NSFG & Data Brief No.~539: Pregnancy Intentions & National Center for Health Statistics & 2022--2023 \\
\midrule
GSS & NSF~25--316: Graduate Students and Postdoctorates in SEH & National Science Foundation & 2023 \\
HERD & NSF~25--313: Higher Education R\&D Expenditures & National Science Foundation & 2023 \\
NSCG & NSF~25--331: National Survey of College Graduates & National Science Foundation & 2023 \\
SDR & NSF~25--320: Survey of Doctorate Recipients & National Science Foundation & 2023 \\
SSERF & NSF~25--318: Survey of S\&E Research Facilities & National Science Foundation & 2023 \\
\midrule
NTEWS & NSF~25--352: National Training, Education, and Workforce Survey & National Center for Education Statistics & 2024 \\
\midrule
NSDUH & 2023 NSDUH Annual National Report & SAMHSA & 2023 \\
\midrule
ASFIN & State Government Finances Summary & U.S. Census Bureau & 2021 \\
ASPEP & Public Employment \& Payroll Summary Report & U.S. Census Bureau & 2024 \\
ASPP & Public Pensions Summary & U.S. Census Bureau & 2019 \\
CPS--ASEC & Income in the United States & U.S. Census Bureau \& BLS & 2024 \\
CPS--ASEC & Poverty in the United States & U.S. Census Bureau \& BLS & 2024 \\
CPS--ASEC & Health Insurance Coverage in the United States & U.S. Census Bureau \& BLS & 2024 \\
RHFS & Rental Housing Finance Survey & U.S. Census Bureau & 2023 \\
STC & State Government Tax Collections Summary Report & U.S. Census Bureau & 2020 \\
\bottomrule
\end{tabular}
}
\end{table*}

\subsection{Survey Abbreviations, Full Names, and Official Links}
\label{app:survey_metadata}

Table~\ref{tab:survey_list} provides the complete list of survey abbreviations used throughout the paper, together with their official full names, releasing agencies, and public data access links. This information is provided to ensure reproducibility and facilitate independent verification of the data sources.

\subsection{Survey--Publication Mapping}
\label{app:survey_pub_mapping}

Table~\ref{tab:survey_pub} summarizes the correspondence between surveys and the official analytical publications from which benchmark queries are extracted. Each query in LongDA is derived from exactly one publication associated with a specific survey, and publications are never exposed to agents during evaluation.

\section{Prompt Design}

\subsection{Prompt Template}
\label{app:prompt_template}
This section presents the complete prompt template used by the ReAct-style agent in all experiments. 
The prompt explicitly specifies the global task description, available tools, question blocks, answer formats, and execution constraints, and is shown verbatim in Figure~\ref{fig:prompt_template} to ensure full reproducibility of the experimental setup.

\subsection{Question Block Example}
\label{app:block_example}
To further illustrate the structure of the task input provided to the agent, Figure~\ref{fig:prompt_example} presents a concrete example of a question block drawn from the \texttt{ASFIN} survey. 
The example contains multiple queries, their required answer formats, and the associated data and documentation files.

\definecolor{myred}{RGB}{220,50,47}
\definecolor{mybrown}{RGB}{110,50,0}

\clearpage
\begin{figure*}[p]
\begin{tcolorbox}[colback=myred!5!white,                
                  colframe=mybrown!80!white,
                  title=Prompt Template]

\textbf{Task:} Analyze the survey data and associated documentation to answer the following questions.

\medskip
\textbf{Survey:} \{\texttt{survey}\}

\medskip
\textbf{Questions and Expected Answer Structures:}
\{\texttt{qa\_block}\}

\medskip
\textbf{Available Data Files:}
\{\texttt{paths/to/the/data/files}\}

\medskip
\textbf{Available Documentation Files:}
\{\texttt{paths/to/the/doc/files}\}

\medskip
\textbf{Instructions:}

\begin{enumerate}
\item \textbf{Understand the Task}
\begin{itemize}
    \item Use the \texttt{prompt} tool to review all questions, expected answer structures, and additional information.
    \item Identify which data files and documentation are required.
\end{itemize}

\item \textbf{Gather Information}
\begin{itemize}
    \item Use \texttt{retriever} or \texttt{search\_doc} to locate relevant documentation.
    \item Use \texttt{read\_doc} to inspect specific files or sections.
    \item Record key information (e.g., variable definitions, column meanings, survey weights) using the \texttt{notes} tool.
    \item Do \emph{not} rely on prior knowledge; always verify from documentation.
\end{itemize}

\item \textbf{Analyze Data}
\begin{itemize}
    \item Write and execute Python code to load and process the data.
    \item Apply sampling weights when required by the survey design.
\end{itemize}

\item \textbf{Format and Save Answers}
\begin{itemize}
    \item Format each answer \emph{exactly} according to the specified answer structure.
    \item Respect unit requirements (e.g., percentages, counts, thousands, millions).
    \item Report all numerical values with \textbf{one decimal place}.
    \item Use the \texttt{answer} tool with \texttt{action='add'} to store each answer by question ID.
    \item Use \texttt{save\_code} to save the analysis code after completing each block.
\end{itemize}

\item \textbf{Working Strategy}
\begin{itemize}
    \item Questions may be solved incrementally; solving all at once is not required.
    \item Use \texttt{answer} with \texttt{action='view'} to inspect saved answers.
    \item Use \texttt{notes} with \texttt{action='list'} to review recorded information.
\end{itemize}
\end{enumerate}

\end{tcolorbox}
\caption{Prompt template used by the agent for solving LongDA tasks.}
\label{fig:prompt_template}
\end{figure*}
\clearpage

\begin{figure*}[p]
\begin{tcolorbox}[colback=myred!5!white,
                  colframe=mybrown!70!white,
                  title=Example Question Block (ASFIN 2021)]

\textbf{Survey:} ASFIN — Annual Survey of State Government Finances (2021)

\medskip
\textbf{Questions and Expected Answer Structures}

\begin{itemize}
\item \textbf{Q1:} What was the total state government revenue in 2021? \\
\textbf{Answer Structure:} \texttt{single\_number} \\
\textbf{Additional Info:} Trillions of dollars.

...

\item \textbf{Q4:} What percentage of total 2021 revenue came from intergovernmental revenue and insurance trust revenue? \\
\textbf{Answer Structure:} \texttt{[intergovernmental\_revenue, insurance\_trust\_revenue]} \\
\textbf{Additional Info:} NA.

...

\item \textbf{Q11:} What was liquor store revenue, utility revenue, miscellaneous general revenue, and current charge revenue in 2021?  \\
\textbf{Answer Structure:} \texttt{[liquor\_store\_revenue, utility\_revenue, miscellaneous\_general\_revenue, current\_charges\_revenue]} \\
\textbf{Additional Info:} Billions of dollars.

...

\end{itemize}

\medskip
\textbf{Available Data Files}
\begin{itemize}
\item \texttt{benchmark/ASFIN/data/21state35.txt}
\end{itemize}

\textbf{Available Documentation Files}
\begin{itemize}
\item \texttt{benchmark/ASFIN/docs/2021 F13\_Form.pdf}
\item \texttt{benchmark/ASFIN/docs/statetechdoc2021.pdf}
\item \texttt{benchmark/ASFIN/docs/public-use-file-layout.xlsx}
\item \texttt{benchmark/ASFIN/docs/itemcodes.xlsx}
\item \texttt{benchmark/ASFIN/docs/government-ids.xlsx}
\end{itemize}

\textbf{Instructions:} {...}

\end{tcolorbox}
\caption{Concrete example of a LongDA question block.}
\label{fig:prompt_example}
\end{figure*}
\clearpage
\section{Case Study: Agent Interaction Example}
\label{app:case_study}

To illustrate how an agent solves LongDA tasks end-to-end, we present a real interaction trace produced by GPT-5 on a block from the NHANES survey (NCHS Data Brief No.~508). 
The block contains multiple related prevalence queries with strict answer-structure constraints. 
Throughout the interaction, the agent has access only to the NHANES microdata and accompanying documentation. 
In this example, GPT-5 successfully answers the entire block.

\subsection{Overview of the Block}
This NHANES block asks for adult (age $\geq 20$) obesity and severe obesity prevalence, both overall and stratified by sex, age group, and education group. 
Solving the block requires the agent to (i) identify the correct BMI variable, (ii) determine the appropriate sampling weight for examination-based analysis, (iii) map demographic variables to subgroup definitions, and (iv) implement weighted estimators in Python while satisfying the required output ordering.

\subsection{Step-by-Step Interaction}

\paragraph{Step 1: Re-establishing task context.}
The agent begins by invoking the \texttt{prompt} tool to re-check the full task specification, including question IDs, answer structures, and auxiliary constraints. 
This behavior is consistent with the strategy we observe for GPT-5 across blocks: it refreshes global task requirements early to minimize downstream formatting errors.

\paragraph{Step 2--4: Documentation navigation and variable identification.}
The primary challenge of this block lies in identifying the correct variables and analytic guidance from long, heterogeneous documentation. 
GPT-5 first uses \texttt{search\_doc} to locate the BMI field in the anthropometry codebook \texttt{BMX\_L.pdf}. 
It identifies \texttt{BMXBMI} and confirms its definition as body mass index (kg/m$^2$), already rounded to one decimal place in the released data. 

In parallel, it queries the demographics codebook \texttt{DEMO\_L.pdf} to locate:
\begin{itemize}
    \item \texttt{RIDAGEYR} for age,
    \item \texttt{RIAGENDR} for sex (coded as 1/2),
    \item \texttt{DMDEDUC2} for adult education categories.
\end{itemize}

The agent then explicitly searches for weight variables in \texttt{DEMO\_L.pdf}, discovering both \texttt{WTMEC2YR} and \texttt{WTINT2YR}, and correctly determines that \texttt{WTMEC2YR} is the appropriate 2-year MEC examination weight for analyses involving examination measures such as BMI. 
This distinction is critical: using interview weights would produce biased estimates.

Finally, GPT-5 consults pregnancy-related documentation and records the discovered variable semantics, coding schemes, subgroup mappings, and weight selection using the \texttt{notes} tool. 
This externalization of key analytical decisions enables clean, reusable code across all queries in the block.

\paragraph{Step 5: Single-pass computation via executable Python.}
After resolving all documentation-dependent uncertainties, GPT-5 performs most of the computation in a single extended code execution. Specifically, it:
\begin{enumerate}
    \item Loads \texttt{DEMO\_L.xpt} and \texttt{BMX\_L.xpt} and merges them on respondent identifier \texttt{SEQN};
    \item Filters to adults aged $\geq 20$ with valid BMI and positive MEC weights (\texttt{WTMEC2YR} $>$ 0);
    \item Constructs indicator variables for obesity (BMI $\geq 30$) and severe obesity (BMI $\geq 40$);
    \item Defines subgroup variables:
    \begin{itemize}
        \item age groups: 20--39, 40--59, 60+;
        \item education groups: \texttt{hs\_or\_less} (DMDEDUC2 $\in \{1,2,3\}$), \texttt{some\_college} (4), \texttt{bachelors\_plus} (5), excluding refused/unknown.
    \end{itemize}
    \item Implements a weighted prevalence estimator
    \[
        \widehat{p} \;=\; 100 \cdot \frac{\sum_i w_i y_i}{\sum_i w_i},
    \]
    using \texttt{WTMEC2YR} as $w_i$ and the indicator variable as $y_i$.
\end{enumerate}

The agent then computes each query by slicing the dataset into the required subpopulation and applying the same estimator. All outputs are rounded to one decimal place to match benchmark formatting constraints.

\paragraph{Answer recording and audit trail.}
Upon completing the computations, GPT-5 records each result using the \texttt{answer} tool under the corresponding question ID, guaranteeing strict adherence to the required output structures. 
It also invokes \texttt{save\_code} to preserve the full analysis pipeline, enabling reproducibility, auditing, and qualitative error analysis.

\paragraph{Step 6: Finalization.}
Finally, the agent emits a structured summary of all answers via \texttt{final\_answer}, serving as a human-readable recap and signaling termination of the interaction. 
Benchmark evaluation, however, relies exclusively on the structured outputs recorded through the \texttt{answer} tool.

\subsection{Implications for LongDA Evaluation}
This example illustrates several core properties that LongDA is explicitly designed to evaluate:
\begin{itemize}
    \item \textbf{Long-document grounding.} Correct solutions require discovering variable definitions, coding schemes, and weight selection rules from documentation rather than relying on prior knowledge alone.
    \item \textbf{Tool-assisted analytical workflow.} Successful execution depends on iterative tool use for information retrieval followed by executable computation.
    \item \textbf{Block-level reuse of intermediate knowledge.} Once variables and weights are identified, the same intermediate understanding supports multiple queries within the block.
    \item \textbf{Strict output compliance.} The agent must satisfy rigid structural and ordering constraints to enable automatic evaluation.
\end{itemize}

\subsection{Summary}
Overall, this interaction exemplifies the intended LongDA workflow: the agent first resolves documentation-dependent uncertainties and then executes a coherent analysis program to answer multiple related questions correctly. 
Such tasks go beyond code synthesis alone and require reliable agentic document navigation and tool use under realistic constraints.


\begingroup
\setlength{\parindent}{0pt}
\tcbset{colback=gray!2,colframe=gray!60,boxrule=0.5pt,arc=2pt,outer arc=2pt,left=6pt,right=6pt,top=4pt,bottom=4pt}
\begin{tcolorbox}[title={Example Trace: GPT-5 on NHANES (Data Brief No.\ 508)},breakable]
\small This appendix shows an execution trace (abridged only by formatting) for a LongDA block. The agent alternates between tool calls (e.g., \texttt{prompt}, \texttt{search\_doc}, \texttt{read\_doc}) and Python code execution, and records final answers via the \texttt{answer} tool.
\end{tcolorbox}

\begin{tcolorbox}[title={Step 1},breakable]
\textbf{Code Action}
\begin{Verbatim}[fontsize=\footnotesize,breaklines=true,breakanywhere=true]
# Step 1: Review the full prompt to confirm tasks
full_prompt = prompt()
print(full_prompt)
\end{Verbatim}
\textbf{Observations}
\begin{Verbatim}[fontsize=\footnotesize,breaklines=true,breakanywhere=true]
Analyze the survey data and relevant documents to answer the following questions.
Survey: NHANES

Questions and Expected Answer Structures:
Question ID: NHANES_NCHS Data Brief No. 508_q1
Q1: What was the overall prevalence of obesity among adults age 20 and older?
Expected Answer Structure: single_number
Additional Info: Body mass index (BMI): Calculated as weight in kilograms divided by height in meters squared, rounded to one decimal place. Obesity: Defined as a BMI of greater than or equal to 30.

Question ID: NHANES_NCHS Data Brief No. 508_q2
Q2: What were the obesity prevalences for men and women?
Expected Answer Structure: [men, women]
Additional Info: NA.

Question ID: NHANES_NCHS Data Brief No. 508_q3
Q3: What were the obesity prevalences for age groups 20-39, 40-59, and 60+?
Expected Answer Structure: [age20_39, age40_59, age60plus]
Additional Info: NA.

Question ID: NHANES_NCHS Data Brief No. 508_q4
Q4: What were the obesity prevalences by education level?
Expected Answer Structure: [hs_or_less, some_college, bachelors_plus]
Additional Info: NA.

Question ID: NHANES_NCHS Data Brief No. 508_q5
Q5: What was the overall prevalence of severe obesity among adults?
Expected Answer Structure: single_number
Additional Info: Severe obesity: Defined as a BMI of greater than or equal to 40.

Question ID: NHANES_NCHS Data Brief No. 508_q6
Q6: What were the severe obesity prevalences for age groups 20-39, 40-59, and 60+?
Expected Answer Structure: [age20_39, age40_59, age60plus]
Additional Info: NA.

Question ID: NHANES_NCHS Data Brief No. 508_q7
Q7: What were the severe obesity prevalences for men and women?
Expected Answer Structure: [men, women]
Additional Info: NA.

Question ID: NHANES_NCHS Data Brief No. 508_q8
Q8: What were the severe obesity prevalences for men by age group?
Expected Answer Structure: [men20_39, men40_59, men60plus]
Additional Info: NA.

Question ID: NHANES_NCHS Data Brief No. 508_q9
Q9: What were the severe obesity prevalences for women by age group?
Expected Answer Structure: [women20_39, women40_59, women60plus]
Additional Info: NA.


Available data files:
[Omitted for simplicity]

Available documentation files:
[Omitted for simplicity]

Instructions:
You are analyzing survey data to answer specific questions. Follow this workflow:

1. UNDERSTAND THE TASK
   - Use `prompt` tool to review all questions, expected answer structures, and additional info
   - Identify which data files and documentation you need

2. GATHER INFORMATION
   - Use `retriever` or `search_doc` to find relevant documentation about the dataset
   - Use `read_doc` to read specific documentation files when needed
   - Record key information (column names, meanings, important variables) using `notes` tool
   - Do NOT rely on your own knowledge - always seek information from the documentation

3. ANALYZE DATA
   - Write Python code to load and analyze the data
   - Apply sampling weights if required by the survey design

4. FORMAT AND SAVE ANSWERS
   - Format your answer EXACTLY as specified in the "Expected Answer Structure"
   - Pay attention to unit requirements (e.g., percentages, counts, thousands, millions, billions)
   - All numbers should have ONE decimal place (e.g., 45.3, not 45 or 45.27)
   - Use `answer` tool with action='add' to save each answer with its question ID
   - Use `save_code` to store the analysis code for each block after finishing the block

5. WORKING STRATEGY
   - You can answer questions one by one - no need to solve all at once
   - Use `answer` tool with action='view' to check answers you've already saved
   - Use `notes` tool with action='list' to review your documented findings
\end{Verbatim}
\end{tcolorbox}
\vspace{4pt}

\begin{tcolorbox}[title={Step 2},breakable]
\textbf{Code Action}
\begin{Verbatim}[fontsize=\footnotesize,breaklines=true,breakanywhere=true]
# Step 2: Gather documentation info: search for BMI and weight variables in BMX and DEMO docs

bmx_doc = "benchmark/NHANES/docs/Examination/BMX_L.pdf"
demo_doc = "benchmark/NHANES/docs/Demographics/DEMO_L.pdf"
exam_overview_doc = "benchmark/NHANES/docs/Examination/August 2021-August 2023 Examination Data - Continuous NHANES.pdf"
demo_overview_doc = "benchmark/NHANES/docs/Demographics/August 2021-August 2023 Demographics Data - Continuous NHANES.pdf"

# Search for BMI variable in BMX_L.pdf
bmi_search = search_doc(doc_path=bmx_doc, keyword="BMXBMI", max_matches=5, context_chars=200)
print("BMX BMI search:\
", bmi_search)

# Search for what BMI means text
bmi_text_search = search_doc(doc_path=bmx_doc, keyword="Body Mass Index", max_matches=5, context_chars=200)
print("\
BMX Body Mass Index search:\
", bmi_text_search)

# Search for exam weights in examination overview (expect WTMEC...)
exam_weight_search = search_doc(doc_path=exam_overview_doc, keyword="weight", max_matches=10, context_chars=200)
print("\
Exam overview weight search:\
", exam_weight_search)

# Search for MEC weight variable name specifically
mec_search = search_doc(doc_path=exam_overview_doc, keyword="WTMEC", max_matches=10, context_chars=200)
print("\
Exam overview WTMEC search:\
", mec_search)

# Search demographic variables: age, sex, education in DEMO_L.pdf
age_search = search_doc(doc_path=demo_doc, keyword="RIDAGEYR", max_matches=5, context_chars=200)
print("\
DEMO age search:\
", age_search)

sex_search = search_doc(doc_path=demo_doc, keyword="RIAGENDR", max_matches=5, context_chars=200)
print("\
DEMO sex search:\
", sex_search)

educ_search = search_doc(doc_path=demo_doc, keyword="DMDEDUC2", max_matches=5, context_chars=250)
print("\
DEMO education search:\
", educ_search)

# Also search the demographics overview for weights guidance
demo_weight_search = search_doc(doc_path=demo_overview_doc, keyword="weight", max_matches=10, context_chars=200)
print("\
Demographics overview weight search:\
", demo_weight_search)
\end{Verbatim}
\textbf{Observations}
\begin{Verbatim}[fontsize=\footnotesize,breaklines=true,breakanywhere=true]
Execution logs:
BMX BMI search:
 [benchmark/NHANES/docs/Examination/BMX_L.pdf] Found 3 matches for 'BMXBMI'.
Page 3: ...Body Mass Index (BMXBMI) : Body Mass Index (BMI) w as calculated as weight in kilogr ams divided b y height in meters squared, and then rounded to one decimal place. BMI Category - Children/Adolescents (BMDBMIC) : This v ar...
Page 14: ...Variable Name: BMXBMI SAS Label: Body Mass Index (kg/m**2) English Text: Body Mass Index (kg/m**2) Target: Both males and females 2 YEARS - 150 YEARSBMXBMI - Body Mass Index (kg/m**2) Code or Value Value Description Count...
Page 14: ...Variable Name: BMXBMI SAS Label: Body Mass Index (kg/m**2) English Text: Body Mass Index (kg/m**2) Target: Both males and females 2 YEARS - 150 YEARSBMXBMI - Body Mass Index (kg/m**2) Code or Value Value Description Count Cumulative Skip to Item 11.1 to 74.8 Range of V alues 8471 8471 . Missing 389 8860...

BMX Body Mass Index search:
 [benchmark/NHANES/docs/Examination/BMX_L.pdf] Found 5 matches for 'Body Mass Index'.
Page 3: ...Body Mass Index (BMXBMI) : Body Mass Index (BMI) w as calculated as weight in kilogr ams divided b y height in meters squared, and then rounded to one decimal place. BMI Category - Children/Adolescents (BMDBMIC) : T...
Page 3: ...Body Mass Index (BMXBMI) : Body Mass Index (BMI) w as calculated as weight in kilogr ams divided b y height in meters squared, and then rounded to one decimal place. BMI Category - Children/Adolescents (BMDBMIC) : This v ariable w as created ...
Page 14: ...Variable Name: BMXBMI SAS Label: Body Mass Index (kg/m**2) English Text: Body Mass Index (kg/m**2) Target: Both males and females 2 YEARS - 150 YEARSBMXBMI - Body Mass Index (kg/m**2) Code or Value Value Description Count Cumulative Skip to Item 11...
Page 14: ...Variable Name: BMXBMI SAS Label: Body Mass Index (kg/m**2) English Text: Body Mass Index (kg/m**2) Target: Both males and females 2 YEARS - 150 YEARSBMXBMI - Body Mass Index (kg/m**2) Code or Value Value Description Count Cumulative Skip to Item 11.1 to 74.8 Range of V alues 8471 8471 . ...
Page 14: ...Variable Name: BMXBMI SAS Label: Body Mass Index (kg/m**2) English Text: Body Mass Index (kg/m**2) Target: Both males and females 2 YEARS - 150 YEARSBMXBMI - Body Mass Index (kg/m**2) Code or Value Value Description Count Cumulative Skip to Item 11.1 to 74.8 Range of V alues 8471 8471 . Missing 389 8860...

Exam overview weight search:
 [benchmark/NHANES/docs/Examination/August 2021-August 2023 Examination Data - Continuous NHANES.pdf] No matches for 'weight'.

Exam overview WTMEC search:
 [benchmark/NHANES/docs/Examination/August 2021-August 2023 Examination Data - Continuous NHANES.pdf] No matches for 'WTMEC'.

DEMO age search:
 [benchmark/NHANES/docs/Demographics/DEMO_L.pdf] Found 4 matches for 'RIDAGEYR'.
Page 2: ...entifies whether a participant w as both interviewed at home or b y telephone and examined in the mobile examination center (MEC), or w as only interviewed but nev er participated in the examination. RIDAGEYR: Age in y ears at the time of the screening interview for surv ey participants between the ages of 1 and 79 y ears of age. Due to disclosure concerns, all responses of participants aged 80 y ears and...
Page 2: ...sure concerns, all responses of participants aged 80 y ears and older are coded as "80. " In NHANES A ugust 2021- August 2023, the weighted mean age for participants 80 y ears and older is 85 y ears. RIDAGEYR w as calculated based on the participant 's date of birth. In r are cases, if the actual date of birth w as missing but the participant 's age in y ears w as pro vided, the reported age w as used. RI...
Page 10: ...Variable Name: RIDAGEYR SAS Label: Age in y ears at screening English Text: Age in y ears of the participant at the time of screening. Individuals 80 and o ver are top-coded at 80 y ears of age. Target: Both males and femal...
Page 10: ...ars at screening English Text: Age in y ears of the participant at the time of screening. Individuals 80 and o ver are top-coded at 80 y ears of age. Target: Both males and females 0 YEARS - 150 YEARSRIDAGEYR - Age in years at screening Code or Value Value Description Count Cumulative Skip to Item 0 to 79 Range of V alues 11408 11408 80 80 years of age and o ver 525 11933 . Missing 0 11933...

DEMO sex search:
 [benchmark/NHANES/docs/Demographics/DEMO_L.pdf] Found 2 matches for 'RIAGENDR'.
Page 9: ...Variable Name: RIAGENDR SAS Label: Gender English Text: Gender of the participant. Target: Both males and females 0 YEARS - 150 YEARSRIAGENDR - Gender Code or Value Value Description Count Cumulative Skip to Item 1 Male 557...
Page 9: ...Variable Name: RIAGENDR SAS Label: Gender English Text: Gender of the participant. Target: Both males and females 0 YEARS - 150 YEARSRIAGENDR - Gender Code or Value Value Description Count Cumulative Skip to Item 1 Male 5575 5575 2 Female 6358 11933 . Missing 0 11933...

DEMO education search:
 [benchmark/NHANES/docs/Demographics/DEMO_L.pdf] Found 5 matches for 'DMDEDUC2'.
Page 3: ...ded into 6 categories: 1 = Less than 1 y ear, 2 = 1 to 4 y ears, 3 = 5 to 9 y ears, 4 = 10 to 14 y ears, 5 = 15 to 19 y ears, and 6 = 20 y ear or more. F or participants who did not report the month of their arriv al, a v alue of 7 (July) w as used. DMDEDUC2: the highest gr ade or lev el of education completed b y adults 20 y ears and older . DMDMARTZ: The marital status question w as ask ed of people 14 y ears of age and older . Due to disclosure risk, marital status is only released for people 20 y ea...
Page 17: ...SAS Label: Country of birth English Text: Country of birth Target: Both males and females 0 YEARS - 150 YEARSDMDBORN4 - Country of birth Code or Value Value Description Count Cumulative Skip to Item 1 Born in 50 US states or W ashington, 10039 10039 DMDEDUC2 2 Others 1875 11914 77 Refused 0 11914 DMDEDUC2 99 Don't know 0 11914 DMDEDUC2 . Missing 19 11933...
Page 17: ...y of birth Target: Both males and females 0 YEARS - 150 YEARSDMDBORN4 - Country of birth Code or Value Value Description Count Cumulative Skip to Item 1 Born in 50 US states or W ashington, 10039 10039 DMDEDUC2 2 Others 1875 11914 77 Refused 0 11914 DMDEDUC2 99 Don't know 0 11914 DMDEDUC2 . Missing 19 11933...
Page 17: ...nd females 0 YEARS - 150 YEARSDMDBORN4 - Country of birth Code or Value Value Description Count Cumulative Skip to Item 1 Born in 50 US states or W ashington, 10039 10039 DMDEDUC2 2 Others 1875 11914 77 Refused 0 11914 DMDEDUC2 99 Don't know 0 11914 DMDEDUC2 . Missing 19 11933...
Page 19: ...Variable Name: DMDEDUC2 SAS Label: Education lev el - Adults 20+ English Text: What is the highest gr ade or lev el of school {y ou ha ve/SP has} completed or the highest degree {y ou ha ve/s/he has} receiv ed? English Instructions: HAND CARD DMQ1 READ HAND CARD CA TEGORIE...

Demographics overview weight search:
 [benchmark/NHANES/docs/Demographics/August 2021-August 2023 Demographics Data - Continuous NHANES.pdf] Found 1 matches for 'weight'.
Page 1: ...rview of Sample Design, Nonresponse Bias Assessment, and Analytic Guidelines SAS Universal Viewer Data User Agreement Data File Name Doc File Data File Date Published Demographic Variables and Sample Weights DEMO_L Doc DEMO_L Data [XPT - 2.5 MB] September 2024...
Last output from code snippet:
None
\end{Verbatim}
\end{tcolorbox}
\vspace{4pt}

\begin{tcolorbox}[title={Step 3},breakable]
\textbf{Code Action}
\begin{Verbatim}[fontsize=\footnotesize,breaklines=true,breakanywhere=true]
# Step 2 (continued): find weight variables in DEMO_L documentation
demo_doc = "benchmark/NHANES/docs/Demographics/DEMO_L.pdf"

wts_mec_search = search_doc(doc_path=demo_doc, keyword="WTMEC", max_matches=10, context_chars=240)
print("DEMO_L search for WTMEC:\
", wts_mec_search)

wts_int_search = search_doc(doc_path=demo_doc, keyword="WTINT", max_matches=10, context_chars=240)
print("\
DEMO_L search for WTINT:\
", wts_int_search)

wts_overall_search = search_doc(doc_path=demo_doc, keyword="weight", max_matches=10, context_chars=240)
print("\
DEMO_L search for \
\end{Verbatim}
\textbf{Observations}
\begin{Verbatim}[fontsize=\footnotesize,breaklines=true,breakanywhere=true]
Execution logs:
DEMO_L search for WTMEC:
 [benchmark/NHANES/docs/Demographics/DEMO_L.pdf] Found 3 matches for 'WTMEC'.
Page 5: ...protect confidentialit y and to reduce disclosure risk and are described in NHANES Analytic Guidelines . Analysts should review the Guidelines carefully prior to analyzing surv ey data. Sample Weights: The 2- year sample weights (WTINT2YR, WTMEC2YR, depending on the analysis) should be used for all NHANES A ugust 2021- August 2023 analyses. Please refer to the NHANES Analytic Guidelines and the on-line NHANES T utorial for further details on the use of sample weights and other ana...
Page 29: ...Variable Name: WTMEC2YR SAS Label: Full sample 2- year MEC exam weight English Text: Full sample 2 y ear MEC exam weight. Target: Both males and females 0 YEARS - 150 YEARSWTMEC2YR - Full sample 2-year MEC exam weight Code or Value Value Description Count Cumu...
Page 29: ...Variable Name: WTMEC2YR SAS Label: Full sample 2- year MEC exam weight English Text: Full sample 2 y ear MEC exam weight. Target: Both males and females 0 YEARS - 150 YEARSWTMEC2YR - Full sample 2-year MEC exam weight Code or Value Value Description Count Cumulative Skip to Item 4581.595095 to 227108.296958 Range of V alues 8860 8860 0 Not MEC Examined 3073 11933 . Missing 0 11933...

DEMO_L search for WTINT:
 [benchmark/NHANES/docs/Demographics/DEMO_L.pdf] Found 3 matches for 'WTINT'.
Page 5: ...e used to protect confidentialit y and to reduce disclosure risk and are described in NHANES Analytic Guidelines . Analysts should review the Guidelines carefully prior to analyzing surv ey data. Sample Weights: The 2- year sample weights (WTINT2YR, WTMEC2YR, depending on the analysis) should be used for all NHANES A ugust 2021- August 2023 analyses. Please refer to the NHANES Analytic Guidelines and the on-line NHANES T utorial for further details on the use of sample weights and...
Page 28: ...Variable Name: WTINT2YR SAS Label: Full sample 2- year interview weight English Text: Full sample 2 y ear interview weight. Target: Both males and females 0 YEARS - 150 YEARSWTINT2YR - Full sample 2-year interview weight Code or Value Value Description Count C...
Page 28: ...Variable Name: WTINT2YR SAS Label: Full sample 2- year interview weight English Text: Full sample 2 y ear interview weight. Target: Both males and females 0 YEARS - 150 YEARSWTINT2YR - Full sample 2-year interview weight Code or Value Value Description Count Cumulative Skip to Item 4584.463196 to 170968.343177 Range of V alues 11933 11933 . Missing 0 11933...

DEMO_L search for 'weight':
 [benchmark/NHANES/docs/Demographics/DEMO_L.pdf] Found 10 matches for 'weight'.
Page 1: ...cific to this cy cle, please see the "Analytic Notes" section below . This demogr aphics file pro vides information on the following topics: Survey participant 's household interview and examination status; Interview and examination sample weights; Masked variance units; The six -month time period when the examination w as performed; Pregnancy status; The ratio of family income to po verty guidelines; Total number of people in the Household; Demogr aphic information about the house...
Page 2: ...of the screening interview for surv ey participants between the ages of 1 and 79 y ears of age. Due to disclosure concerns, all responses of participants aged 80 y ears and older are coded as "80. " In NHANES A ugust 2021- August 2023, the weighted mean age for participants 80 y ears and older is 85 y ears. RIDAGEYR w as calculated based on the participant 's date of birth. In r are cases, if the actual date of birth w as missing but the participant 's age in y ears w as pro vided,...
Page 5: ...sample design variance units. MVUs are used to protect confidentialit y and to reduce disclosure risk and are described in NHANES Analytic Guidelines . Analysts should review the Guidelines carefully prior to analyzing surv ey data. Sample Weights: The 2- year sample weights (WTINT2YR, WTMEC2YR, depending on the analysis) should be used for all NHANES A ugust 2021- August 2023 analyses. Please refer to the NHANES Analytic Guidelines and the on-line NHANES T utorial for further deta...
Page 5: .... MVUs are used to protect confidentialit y and to reduce disclosure risk and are described in NHANES Analytic Guidelines . Analysts should review the Guidelines carefully prior to analyzing surv ey data. Sample Weights: The 2- year sample weights (WTINT2YR, WTMEC2YR, depending on the analysis) should be used for all NHANES A ugust 2021- August 2023 analyses. Please refer to the NHANES Analytic Guidelines and the on-line NHANES T utorial for further details on the use of sample wei...
Page 5: ...ghts (WTINT2YR, WTMEC2YR, depending on the analysis) should be used for all NHANES A ugust 2021- August 2023 analyses. Please refer to the NHANES Analytic Guidelines and the on-line NHANES T utorial for further details on the use of sample weights and other analytic issues. Disclosure risks and issues pertaining to confidentialit y protection prev ent NCHS from releasing all the NHANES demogr aphic v ariables publicly . Additional information and instructions for requesting use of ...
Page 28: ...Variable Name: WTINT2YR SAS Label: Full sample 2- year interview weight English Text: Full sample 2 y ear interview weight. Target: Both males and females 0 YEARS - 150 YEARSWTINT2YR - Full sample 2-year interview weight Code or Value Value Description Count Cumulative Skip to Item 4584.463196 to 170968.343177...
Page 28: ...Variable Name: WTINT2YR SAS Label: Full sample 2- year interview weight English Text: Full sample 2 y ear interview weight. Target: Both males and females 0 YEARS - 150 YEARSWTINT2YR - Full sample 2-year interview weight Code or Value Value Description Count Cumulative Skip to Item 4584.463196 to 170968.343177 Range of V alues 11933 11933 . Missing 0 11933...
Page 28: ...Variable Name: WTINT2YR SAS Label: Full sample 2- year interview weight English Text: Full sample 2 y ear interview weight. Target: Both males and females 0 YEARS - 150 YEARSWTINT2YR - Full sample 2-year interview weight Code or Value Value Description Count Cumulative Skip to Item 4584.463196 to 170968.343177 Range of V alues 11933 11933 . Missing 0 11933...
Page 29: ...Variable Name: WTMEC2YR SAS Label: Full sample 2- year MEC exam weight English Text: Full sample 2 y ear MEC exam weight. Target: Both males and females 0 YEARS - 150 YEARSWTMEC2YR - Full sample 2-year MEC exam weight Code or Value Value Description Count Cumulative Skip to Item 4581.595095 to 227108.296958 R...
Page 29: ...Variable Name: WTMEC2YR SAS Label: Full sample 2- year MEC exam weight English Text: Full sample 2 y ear MEC exam weight. Target: Both males and females 0 YEARS - 150 YEARSWTMEC2YR - Full sample 2-year MEC exam weight Code or Value Value Description Count Cumulative Skip to Item 4581.595095 to 227108.296958 Range of V alues 8860 8860 0 Not MEC Examined 3073 ...

DEMO_L page 19 (likely DMDEDUC2 details):
 [benchmark/NHANES/docs/Demographics/DEMO_L.pdf] Pages: Page 19
Page 19: Variable Name: DMDEDUC2 SAS Label: Education lev el - Adults 20+ English Text: What is the highest gr ade or lev el of school {y ou ha ve/SP has} completed or the highest degree {y ou ha ve/s/he has} receiv ed? English Instructions: HAND CARD DMQ1 READ HAND CARD CA TEGORIES IF NECES SARY ENTER HIGHEST LEVEL OF SCHOOL Target: Both males and females 20 YEARS - 150 YEARSDMDEDUC2 - Education level - Adults 20+ Code or ValueValue Description Count Cumulative Skip to Item 1 Less than 9th gr ade 373 373 2 9-11th gr ade (Includes 12th gr ade with no diploma)666 1039 3 High school gr aduate/GED or equiv alent 1749 2788 4 Some college or AA degree 2370 5158 5 College gr aduate or abo ve 2625 7783 7 Refused 0 7783 9 Don't know 11 7794 . Missing 4139 11933
Last output from code snippet:
None
\end{Verbatim}
\end{tcolorbox}
\vspace{4pt}

\begin{tcolorbox}[title={Step 4},breakable]
\textbf{Code Action}
\begin{Verbatim}[fontsize=\footnotesize,breaklines=true,breakanywhere=true]
# Step 2 (additional): check for pregnancy variable in DEMO_L to see if exclusions are required
demo_doc = "benchmark/NHANES/docs/Demographics/DEMO_L.pdf"
preg_search = search_doc(doc_path=demo_doc, keyword="preg", max_matches=10, context_chars=200)
print("DEMO_L pregnancy search:\
", preg_search)

# Record key notes about variables and weights
notes(action="add", content="Use BMX_L for BMI: BMXBMI is Body Mass Index (kg/m**2), rounded to one decimal. Use DEMO_L for demographics: RIDAGEYR (age in years), RIAGENDR (1=Male, 2=Female), DMDEDUC2 (education for adults 20+). Use WTMEC2YR for exam-based analyses per DEMO_L (2-year MEC exam weight).", index=0)
notes(action="add", content="Education grouping for adults 20+: hs_or_less = DMDEDUC2 in [1,2,3]; some_college = 4; bachelors_plus = 5. Exclude DMDEDUC2 in [7,9] or missing for education-specific estimates.", index=0)
notes(action="add", content="Age groups: 20-39, 40-59, 60+ using RIDAGEYR (80+ top-coded to 80, acceptable).", index=0)
print(notes(action="list", content="", index=0))
\end{Verbatim}
\textbf{Observations}
\begin{Verbatim}[fontsize=\footnotesize,breaklines=true,breakanywhere=true]
Execution logs:
DEMO_L pregnancy search:
 [benchmark/NHANES/docs/Demographics/DEMO_L.pdf] Found 10 matches for 'preg'.
Page 1: ...ics: Survey participant 's household interview and examination status; Interview and examination sample weights; Masked variance units; The six -month time period when the examination w as performed; Pregnancy status; The ratio of family income to po verty guidelines; Total number of people in the Household; Demogr aphic information about the household reference person; and Other selected demogr aphic...
Page 3: ...d for people 20 y ears of age and older , and recoded from the original 6 categories to 3 categories (1 = Married/Living with partner; 2 = Widowed/Div orced/Separ ated; 3 = Nev er married). RIDEXPRG: Pregnancy status at the time of the health examination w as ascertained for females 8-59 y ears of age. Due to disclosure risk, pregnancy status is released only for women 20-44 y ears of age. RIDEXPRG v ...
Page 3: ...rtner; 2 = Widowed/Div orced/Separ ated; 3 = Nev er married). RIDEXPRG: Pregnancy status at the time of the health examination w as ascertained for females 8-59 y ears of age. Due to disclosure risk, pregnancy status is released only for women 20-44 y ears of age. RIDEXPRG v alues are based on self -reported pregnancy status and urine pregnancy test results. P eople who reported they were pregnant at ...
Page 3: ...alth examination w as ascertained for females 8-59 y ears of age. Due to disclosure risk, pregnancy status is released only for women 20-44 y ears of age. RIDEXPRG v alues are based on self -reported pregnancy status and urine pregnancy test results. P eople who reported they were pregnant at the time of exam were assumed to be pregnant (RIDEXPRG=1). Those who reported they were not pregnant or did no...
Page 3: ...tained for females 8-59 y ears of age. Due to disclosure risk, pregnancy status is released only for women 20-44 y ears of age. RIDEXPRG v alues are based on self -reported pregnancy status and urine pregnancy test results. P eople who reported they were pregnant at the time of exam were assumed to be pregnant (RIDEXPRG=1). Those who reported they were not pregnant or did not know their pregnancy stat...
Page 3: ...e risk, pregnancy status is released only for women 20-44 y ears of age. RIDEXPRG v alues are based on self -reported pregnancy status and urine pregnancy test results. P eople who reported they were pregnant at the time of exam were assumed to be pregnant (RIDEXPRG=1). Those who reported they were not pregnant or did not know their pregnancy status were further classified based on the results of the ...
Page 3: ...men 20-44 y ears of age. RIDEXPRG v alues are based on self -reported pregnancy status and urine pregnancy test results. P eople who reported they were pregnant at the time of exam were assumed to be pregnant (RIDEXPRG=1). Those who reported they were not pregnant or did not know their pregnancy status were further classified based on the results of the urine pregnancy test. If the respondent reported...
Page 3: ...elf -reported pregnancy status and urine pregnancy test results. P eople who reported they were pregnant at the time of exam were assumed to be pregnant (RIDEXPRG=1). Those who reported they were not pregnant or did not know their pregnancy status were further classified based on the results of the urine pregnancy test. If the respondent reported "no" or "don' t know" and the urine test result w as po...
Page 3: ...and urine pregnancy test results. P eople who reported they were pregnant at the time of exam were assumed to be pregnant (RIDEXPRG=1). Those who reported they were not pregnant or did not know their pregnancy status were further classified based on the results of the urine pregnancy test. If the respondent reported "no" or "don' t know" and the urine test result w as positiv e, the respondent w as co...
Page 3: ...t the time of exam were assumed to be pregnant (RIDEXPRG=1). Those who reported they were not pregnant or did not know their pregnancy status were further classified based on the results of the urine pregnancy test. If the respondent reported "no" or "don' t know" and the urine test result w as positiv e, the respondent w as coded as pregnant (RIDEXPRG=1). If the respondent reported "no" and the urine...
Notes:
1. Use BMX_L for BMI: BMXBMI is Body Mass Index (kg/m**2), rounded to one decimal. Use DEMO_L for demog...
2. Education grouping for adults 20+: hs_or_less = DMDEDUC2 in [1,2,3]; some_college = 4; bachelors_plu...
3. Age groups: 20-39, 40-59, 60+ using RIDAGEYR (80+ top-coded to 80, acceptable).
Last output from code snippet:
None
\end{Verbatim}
\end{tcolorbox}
\vspace{4pt}

\begin{tcolorbox}[title={Step 5},breakable]
\textbf{Code Action}
\begin{Verbatim}[fontsize=\footnotesize,breaklines=true,breakanywhere=true]
# Check BMX documentation for any mention of excluding pregnant women
bmx_doc = "benchmark/NHANES/docs/Examination/BMX_L.pdf"
preg_search_bmx = search_doc(doc_path=bmx_doc, keyword="preg", max_matches=10, context_chars=200)
print("BMX_L pregnancy search:\
", preg_search_bmx)

# Proceed to data analysis
import pandas as pd
import numpy as np

# Load data
demo_path = "benchmark/NHANES/data/Demographics/DEMO_L.xpt"
bmx_path = "benchmark/NHANES/data/Examination/BMX_L.xpt"

demo = pd.read_sas(demo_path, format="xport")
bmx = pd.read_sas(bmx_path, format="xport")

# Keep only necessary columns to reduce memory
demo_cols = ["SEQN", "RIDAGEYR", "RIAGENDR", "DMDEDUC2", "WTMEC2YR"]
demo = demo[demo_cols]
bmx_cols = ["SEQN", "BMXBMI"]
bmx = bmx[bmx_cols]

# Merge
df = demo.merge(bmx, on="SEQN", how="inner")

# Convert to numeric (ensure dtypes)
for col in ["RIDAGEYR", "RIAGENDR", "DMDEDUC2", "WTMEC2YR", "BMXBMI"]:
    df[col] = pd.to_numeric(df[col], errors="coerce")

# Restrict to adults 20+ and with valid weight and BMI
adults = df[(df["RIDAGEYR"] >= 20) & (df["WTMEC2YR"] > 0) & df["BMXBMI"].notna()]

# Define obesity and severe obesity
adults = adults.copy()
adults["obese"] = (adults["BMXBMI"] >= 30).astype(int)
adults["severe_obese"] = (adults["BMXBMI"] >= 40).astype(int)

# Age groups
def age_group(age):
    if age >= 20 and age <= 39:
        return "20-39"
    elif age >= 40 and age <= 59:
        return "40-59"
    else:
        return "60+"
adults["age_group"] = adults["RIDAGEYR"].apply(age_group)

# Education groups: hs_or_less=[1,2,3], some_college=4, bachelors_plus=5
def educ_group(x):
    if x in [1, 2, 3]:
        return "hs_or_less"
    elif x == 4:
        return "some_college"
    elif x == 5:
        return "bachelors_plus"
    else:
        return np.nan
adults["educ_group"] = adults["DMDEDUC2"].apply(educ_group)

# Helper to compute weighted prevalence (%)
def weighted_pct(data, indicator_col, weight_col="WTMEC2YR"):
    w = data[weight_col]
    y = data[indicator_col]
    num = (w * y).sum()
    den = w.sum()
    if den == 0 or np.isnan(den):
        return np.nan
    return float(100.0 * num / den)

# Q1: overall prevalence of obesity among adults 20+
q1_val = weighted_pct(adults, "obese")
q1 = round(q1_val, 1) if pd.notna(q1_val) else np.nan
print("Q1 overall obesity prevalence:", q1)

# Q2: obesity prevalences for men and women [men, women]
men = adults[adults["RIAGENDR"] == 1]
women = adults[adults["RIAGENDR"] == 2]
q2_vals = [round(weighted_pct(men, "obese"), 1), round(weighted_pct(women, "obese"), 1)]
print("Q2 obesity men, women:", q2_vals)

# Q3: obesity by age groups [20-39, 40-59, 60+]
q3_vals = []
for grp in ["20-39", "40-59", "60+"]:
    d = adults[adults["age_group"] == grp]
    q3_vals.append(round(weighted_pct(d, "obese"), 1))
print("Q3 obesity by age groups:", q3_vals)

# Q4: obesity by education level [hs_or_less, some_college, bachelors_plus]
educ_order = ["hs_or_less", "some_college", "bachelors_plus"]
q4_vals = []
for eg in educ_order:
    d = adults[adults["educ_group"] == eg]
    q4_vals.append(round(weighted_pct(d, "obese"), 1))
print("Q4 obesity by education level:", q4_vals)

# Q5: overall prevalence of severe obesity among adults
q5_val = weighted_pct(adults, "severe_obese")
q5 = round(q5_val, 1) if pd.notna(q5_val) else np.nan
print("Q5 overall severe obesity prevalence:", q5)

# Q6: severe obesity by age groups [20-39, 40-59, 60+]
q6_vals = []
for grp in ["20-39", "40-59", "60+"]:
    d = adults[adults["age_group"] == grp]
    q6_vals.append(round(weighted_pct(d, "severe_obese"), 1))
print("Q6 severe obesity by age groups:", q6_vals)

# Q7: severe obesity prevalences for men and women [men, women]
q7_vals = [round(weighted_pct(men, "severe_obese"), 1), round(weighted_pct(women, "severe_obese"), 1)]
print("Q7 severe obesity men, women:", q7_vals)

# Q8: severe obesity prevalences for men by age group [men20-39, men40-59, men60+]
q8_vals = []
for grp in ["20-39", "40-59", "60+"]:
    d = men[men["age_group"] == grp]
    q8_vals.append(round(weighted_pct(d, "severe_obese"), 1))
print("Q8 severe obesity men by age:", q8_vals)

# Q9: severe obesity prevalences for women by age group [women20-39, women40-59, women60+]
q9_vals = []
for grp in ["20-39", "40-59", "60+"]:
    d = women[women["age_group"] == grp]
    q9_vals.append(round(weighted_pct(d, "severe_obese"), 1))
print("Q9 severe obesity women by age:", q9_vals)

# Save answers
ans1 = answer(action="add", q_id="NHANES_NCHS Data Brief No. 508_q1", answer=str(q1))
ans2 = answer(action="add", q_id="NHANES_NCHS Data Brief No. 508_q2", answer=[q2_vals[0], q2_vals[1]])
ans3 = answer(action="add", q_id="NHANES_NCHS Data Brief No. 508_q3", answer=[q3_vals[0], q3_vals[1], q3_vals[2]])
ans4 = answer(action="add", q_id="NHANES_NCHS Data Brief No. 508_q4", answer=[q4_vals[0], q4_vals[1], q4_vals[2]])
ans5 = answer(action="add", q_id="NHANES_NCHS Data Brief No. 508_q5", answer=str(q5))
ans6 = answer(action="add", q_id="NHANES_NCHS Data Brief No. 508_q6", answer=[q6_vals[0], q6_vals[1], q6_vals[2]])
ans7 = answer(action="add", q_id="NHANES_NCHS Data Brief No. 508_q7", answer=[q7_vals[0], q7_vals[1]])
ans8 = answer(action="add", q_id="NHANES_NCHS Data Brief No. 508_q8", answer=[q8_vals[0], q8_vals[1], q8_vals[2]])
ans9 = answer(action="add", q_id="NHANES_NCHS Data Brief No. 508_q9", answer=[q9_vals[0], q9_vals[1], q9_vals[2]])

print(ans1, ans2, ans3, ans4, ans5, ans6, ans7, ans8, ans9)

# Save the analysis code
save_code("""
Loaded DEMO_L and BMX_L, merged by SEQN. Filtered adults 20+ with valid BMI and WTMEC2YR > 0.
Defined obesity (BMXBMI >= 30) and severe obesity (BMXBMI >= 40).
Constructed age groups (20-39, 40-59, 60+) and education groups (hs_or_less=[1,2,3], some_college=4, bachelors_plus=5).
Computed weighted prevalences using WTMEC2YR for overall and subgroups, expressed as percentages with one decimal place.
Recorded answers using the answer tool.
""")
\end{Verbatim}
\textbf{Observations}
\begin{Verbatim}[fontsize=\footnotesize,breaklines=true,breakanywhere=true]
Execution logs:
BMX_L pregnancy search:
 [benchmark/NHANES/docs/Examination/BMX_L.pdf] Found 3 matches for 'preg'.
Page 2: ... considered identifiable and pose a disclosure concern. The body weight data for participants who had limb amputations were set to "missing. " This data file includes body measures for women who were pregnant at the time of their health examination. Pregnancy status at the time of the health examination is indicated b y the variable, RIDEXPRG, in the Demogr aphic data file. RIDEXPRG v alues are report...
Page 2: ...cern. The body weight data for participants who had limb amputations were set to "missing. " This data file includes body measures for women who were pregnant at the time of their health examination. Pregnancy status at the time of the health examination is indicated b y the variable, RIDEXPRG, in the Demogr aphic data file. RIDEXPRG v alues are reported for women 20-44 y ears of age. RIDEXPRG for sev...
Page 2: ...status at the time of the health examination is indicated b y the variable, RIDEXPRG, in the Demogr aphic data file. RIDEXPRG v alues are reported for women 20-44 y ears of age. RIDEXPRG for sev eral pregnant women who were outside of this age range are not reported due to disclosure concerns. The body measures data for these participants are also not reported. The NHANES Anthropometry Procedures Manu...
Q1 overall obesity prevalence: 40.3
Q2 obesity men, women: [39.2, 41.3]
Q3 obesity by age groups: [35.6, 46.4, 38.9]
Q4 obesity by education level: [44.5, 45.2, 31.6]
Q5 overall severe obesity prevalence: 9.4
Q6 severe obesity by age groups: [9.6, 11.9, 6.6]
Q7 severe obesity men, women: [6.7, 12.1]
Q8 severe obesity men by age: [6.1, 9.2, 4.3]
Q9 severe obesity women by age: [13.0, 14.6, 8.4]
Answer recorded for question NHANES_NCHS Data Brief No. 508_q1 Answer recorded for question NHANES_NCHS Data Brief No. 508_q2 Answer recorded for question NHANES_NCHS Data Brief No. 508_q3 Answer recorded for question NHANES_NCHS Data Brief No. 508_q4 Answer recorded for question NHANES_NCHS Data Brief No. 508_q5 Answer recorded for question NHANES_NCHS Data Brief No. 508_q6 Answer recorded for question NHANES_NCHS Data Brief No. 508_q7 Answer recorded for question NHANES_NCHS Data Brief No. 508_q8 Answer recorded for question NHANES_NCHS Data Brief No. 508_q9
Last output from code snippet:
Code stored for block NHANES_NCHS Data Brief No. 508.
\end{Verbatim}
\end{tcolorbox}
\vspace{4pt}

\begin{tcolorbox}[title={Step 6},breakable]
\textbf{Code Action}
\begin{Verbatim}[fontsize=\footnotesize,breaklines=true,breakanywhere=true]
summary = {
    "NHANES_NCHS Data Brief No. 508_q1": q1,
    "NHANES_NCHS Data Brief No. 508_q2": q2_vals,
    "NHANES_NCHS Data Brief No. 508_q3": q3_vals,
    "NHANES_NCHS Data Brief No. 508_q4": q4_vals,
    "NHANES_NCHS Data Brief No. 508_q5": q5,
    "NHANES_NCHS Data Brief No. 508_q6": q6_vals,
    "NHANES_NCHS Data Brief No. 508_q7": q7_vals,
    "NHANES_NCHS Data Brief No. 508_q8": q8_vals,
    "NHANES_NCHS Data Brief No. 508_q9": q9_vals
}
final_answer(summary)
\end{Verbatim}
\textbf{Observations}
\begin{Verbatim}[fontsize=\footnotesize,breaklines=true,breakanywhere=true]
Execution logs:
Last output from code snippet:
{'NHANES_NCHS Data Brief No. 508_q1': 40.3, 'NHANES_NCHS Data Brief No. 508_q2': [39.2, 41.3], 'NHANES_NCHS Data Brief No. 508_q3': [35.6, 46.4, 38.9], 'NHANES_NCHS Data Brief No. 508_q4': [44.5, 45.2, 31.6], 'NHANES_NCHS Data Brief No. 508_q5': 9.4, 'NHANES_NCHS Data Brief No. 508_q6': [9.6, 11.9, 6.6], 'NHANES_NCHS Data Brief No. 508_q7': [6.7, 12.1], 'NHANES_NCHS Data Brief No. 508_q8': [6.1, 9.2, 4.3], 'NHANES_NCHS Data Brief No. 508_q9': [13.0, 14.6, 8.4]}
\end{Verbatim}
\end{tcolorbox}
\vspace{4pt}

\endgroup

\end{document}